\documentclass[10pt,journal,compsoc]{IEEEtran}

\ifCLASSOPTIONcompsoc
  \usepackage[nocompress]{cite}
\else
  \usepackage{cite}
\fi

\usepackage{booktabs} 
\usepackage{url}
\usepackage{hyperref}
\usepackage{balance} 
\usepackage[ruled]{algorithm2e} 
\usepackage{graphicx}
\usepackage{graphicx, subfigure, epsfig, multirow, bbding}
\usepackage{booktabs}
\usepackage{array}
\usepackage{caption}
\usepackage{fancynum}
\usepackage{listings}
\usepackage{color}

\definecolor{mygreen}{rgb}{0,0.6,0}
\definecolor{mygray}{rgb}{0.5,0.5,0.5}
\definecolor{mymauve}{rgb}{0.58,0,0.82}

\SetAlFnt{\small}
\SetAlCapFnt{\small}
\SetAlCapNameFnt{\small}
\SetAlCapHSkip{0pt}
\IncMargin{-\parindent}

\newcommand{\code}[1]{{\fontfamily{cmtt}\fontseries{m}\fontshape{n}\selectfont\small{#1}}}

\newcommand{\tab}{\hspace*{1em}}

\ifCLASSINFOpdf
\else
\fi
\begin{document}

\title{Dating with Scambots: Understanding the Ecosystem of Fraudulent Dating Applications}

\author{Yangyu Hu, Haoyu Wang,~\IEEEmembership{Member,~IEEE,} Yajin Zhou,~\IEEEmembership{Member,~IEEE,} Yao Guo,~\IEEEmembership{Member,~IEEE,}\\
Li Li,~\IEEEmembership{Member,~IEEE,} 
Bingxuan Luo and~Fangren Xu 
\IEEEcompsocitemizethanks{\IEEEcompsocthanksitem Yangyu Hu, Haoyu Wang and Bingxuan Luo are with Beijing University of Posts and Telecommunications, China.
E-mail: \{huyangyu910731, haoyuwang, unique\_girl\}@bupt.edu.cn
\IEEEcompsocthanksitem Yajin Zhou is with Zhejiang University. Email: yajin\_zhou@zju.edu.cn
\IEEEcompsocthanksitem Yao Guo is with Peking University. Email: yaoguo@pku.edu.cn
\IEEEcompsocthanksitem Li Li is with Monash University.
Email: li.li@monash.edu
\IEEEcompsocthanksitem Fangren Xu is with Rivermont Collegiate.
Email:fangrenx@rvmt.org
\IEEEcompsocthanksitem Haoyu Wang and Yajin Zhou are co-corresponding authors.}

}

\IEEEtitleabstractindextext{%
\begin{abstract}
In this work, we are focusing on a new and yet uncovered way for
malicious apps to gain profit. They claim to be dating apps. However, their sole purpose is to lure users into purchasing premium/VIP services to start conversations with other (likely fake female) accounts in the app. We call these apps as \emph{fraudulent dating apps}.\\
This paper performs a systematic study to understand the whole ecosystem of fraudulent dating apps. 
Specifically, we have proposed a three-phase method to detect them and subsequently comprehend their characteristics via analyzing the existing account profiles. Our observation reveals that most of the accounts are not managed by real persons, but by chatbots based on predefined conversation templates.
We also analyze the business model of these apps and reveal that multiple parties are actually involved in the ecosystem, including \emph{producers} who develop apps, \emph{publishers} who publish apps
to gain profit, and \emph{the distribution network} that is responsible for distributing apps to end users. Finally, we analyze the impact of them to users (i.e., victims) and estimate the overall revenue.
Our work is the first systematic study on fraudulent dating apps, and the results demonstrate the urge for a solution to protect users.

\end{abstract}

\begin{IEEEkeywords}
Fraud, Mobile App, Dating App, Malware, Android.
\end{IEEEkeywords}}

\maketitle

\IEEEdisplaynontitleabstractindextext

\IEEEpeerreviewmaketitle

\IEEEraisesectionheading{\section{Introduction}\label{sec:introduction}}


\IEEEPARstart{M}{obile} malware is rapidly becoming a serious threat in recent years. The main incentive for attackers to
develop malware is that they could gain illegal profit. For example, previous research
showed that malware authors could gain a profit by injecting advertisements
in benign applications (or apps in short)~\cite{PiggyApp}, or by sending SMS messages to premium-rate numbers~\cite{Sok}. With the deployment of new defenses in the latest Android versions, these methods become less effective.
However, we have observed a trend that malware authors have invented new ways to make a profit.

In this paper, we focus on a new and yet uncovered way for malicious apps to make a profit. Users are lured into installing a
particular kind of dating apps and paying subscription fees for the right to chatting with existing users. However, the sole purpose of these apps is to cheat new users into paying, as the existing accounts in these apps are usually fake identities managed by chatbots. This kind of apps is therefore referenced as \emph{fraudulent dating apps} (or FD apps in short).

\smallskip
\noindent{\bf Properties of FD Apps}\tab
FD apps are usually distributed through online advertising networks, enticing users to install them with attractive pictures or fake claims. Once it is installed, users need to register an account to use the app. Surprisingly,
this process is much simpler than what we have expected. 
Indeed, during our analysis of some apps,
we find that the user only needs to click a few buttons to register a new account, without providing any personal details such as email address or phone number. This is different from the
traditional malware that aims to steal user's private information. 

\begin{figure*}[t]
	\centering
	\subfigure[Many users want to start conversation within a few minutes]{
		\label{subfig:example1}
		\includegraphics[width=0.27\textwidth]{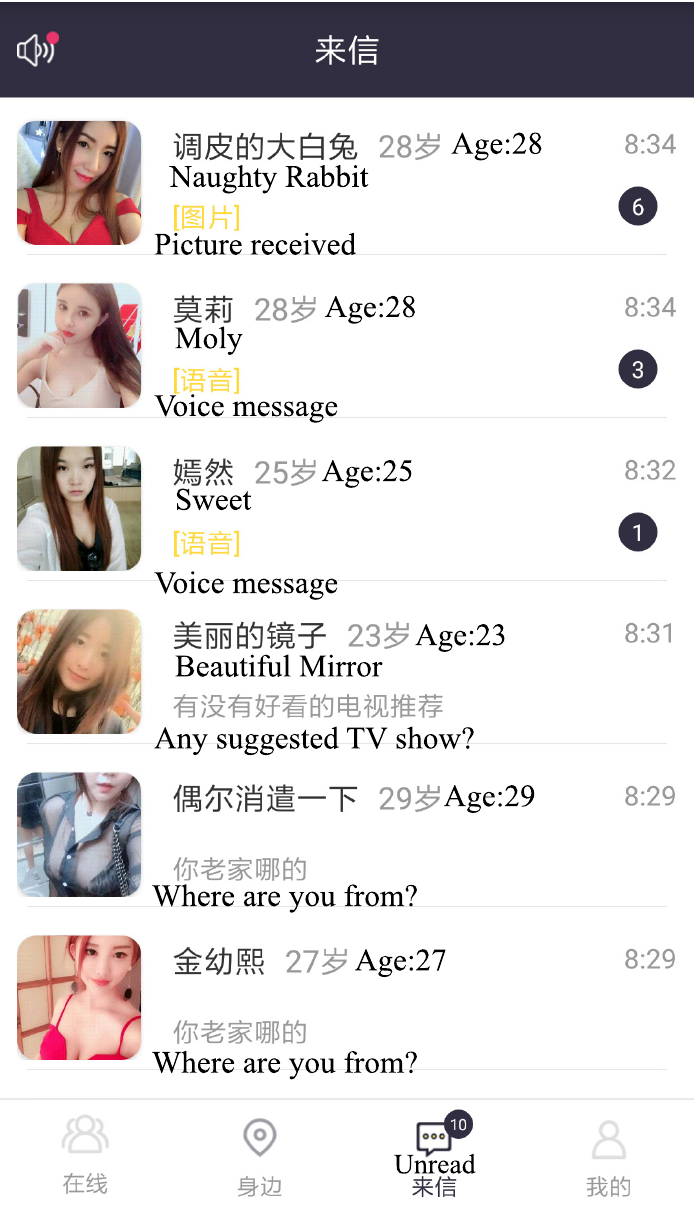}}
	\hspace{0.5ex}
	\subfigure[The messages are usually irrelevant to the conversation]{
		\label{subfig:example2}
		\includegraphics[width=0.27\textwidth]{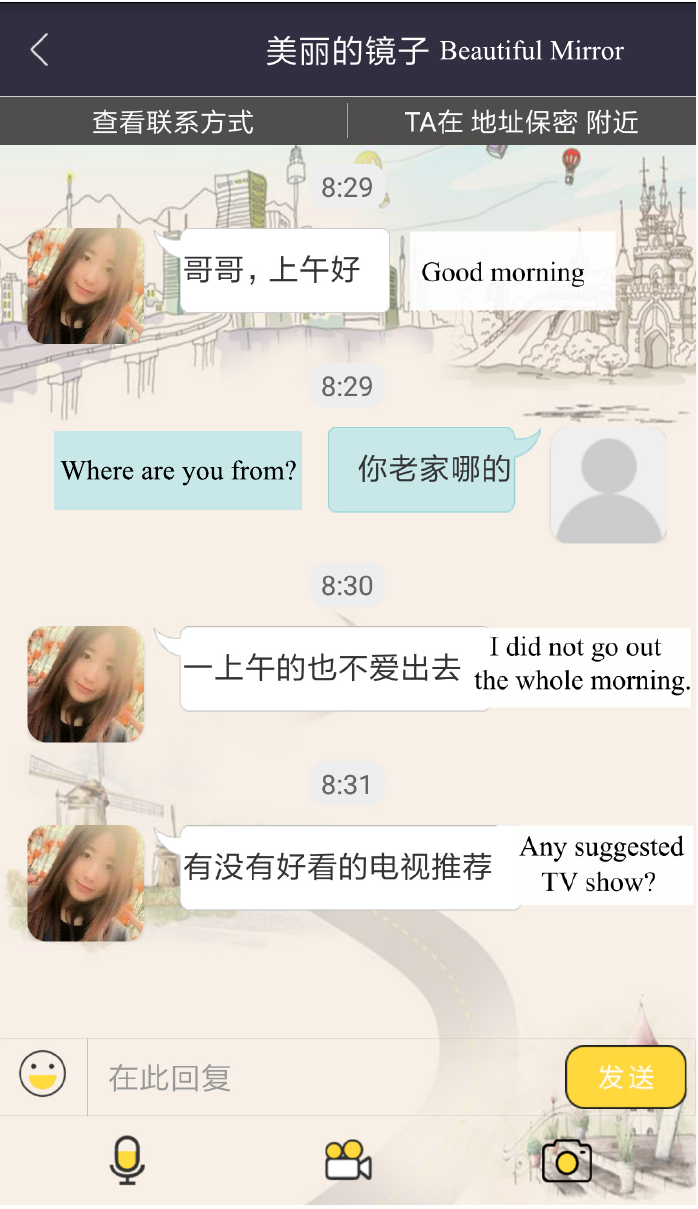}}
	\hspace{0.5ex}
	\subfigure[Premium service is needed to continue conversation]{
		\label{subfig:example3}
		\includegraphics[width=0.27\textwidth]{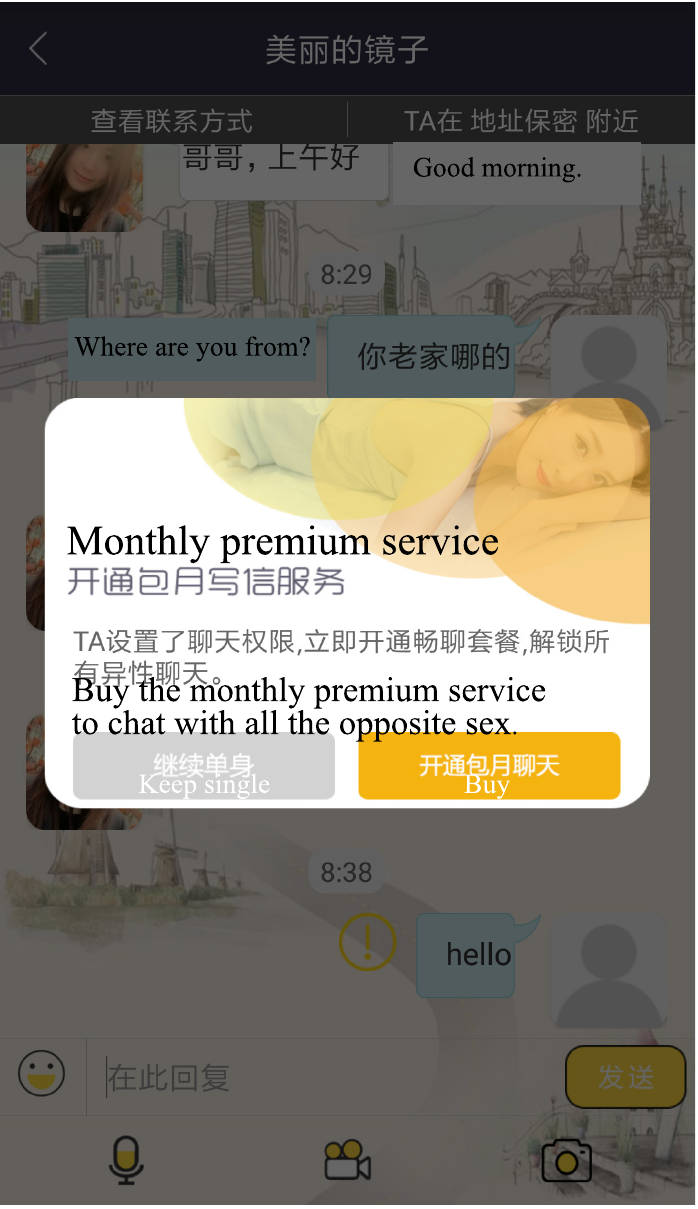}}
	\caption{An example of a FD app.}
	\label{fig:example}
\end{figure*}

After registration and logging into
the app, many (female) accounts will initialize conversation requests to the user within a few minutes
and likely with seductive words or pictures. For instance, during the manual analysis of one of the apps in our study,
seven users sent conversation requests after logging into the app for 5 minutes~(Figure~\ref{subfig:example1}).
All of them were female with attractive profile avatars. Two different users (the last two)
were sending the same message (``Where are you from?'') at the same time (8:29).

Due to this abnormal user behavior, we suspect that the existing accounts in the app are not real people but
chatbots. To confirm our speculation, we sent messages to a randomly picked user (nickname: Beautiful Mirror),
and the replied messages were irrelevant to
the topic of the conversation~(Figure~\ref{subfig:example2}).
Interestingly, after sending one message, we cannot send messages for free anymore.

In order to continue the conversation, we have to subscribe to the
monthly premium service~(Figure~\ref{subfig:example3}) with a cost around seven US dollars.
After purchasing the service, the previously communicated users stop responding
immediately and there were no more users attempting to communicate with us anymore.

This app is not a single case: there exists an underground ecosystem for these FD apps. 
In particular, if a dating app shows the following behaviors,
it can be categorized as a FD app. 1) \emph{abnormal user behaviors}:
after a new user logs into the app, several users will start a conversation in a short period, e.g., a few minutes;
2) \emph{irrelevant messages}: messages are usually irrelevant to the conversation and
out of the topic; 3) \emph{premium services}:
a new user cannot send or can only send a few messages for free,
unless a premium service is purchased. Finally, after purchasing the service,
other users suddenly stop responding to any messages. 
To distinguish with the fake accounts in the app, we call
the new user who is purchasing the services as a \emph{victim}.

\smallskip
\noindent{\bf Study Overview}\tab 
In this paper, we perform a systematic study of FD apps,
including the characterization of the user profiles and interaction patterns,
the business model and involved parties in the ecosystem, as well as the distribution and the impact of these apps to victims.
In particular, our research aims to answer the following questions.
{First}, are the existing user accounts in FD apps are real persons or chatbots? 
{Second}, what is the business model of the FD apps, and
what parties are involved in the ecosystem?
{Third}, how are the FD apps distributed?
{Fourth}, what is the impact of these FD apps to mobile users? For instance, how much money might be charged to a victim?

To this end, we first propose a method to detect FD apps
from 2.5 million apps downloaded from nine third-party Android app
markets\footnote{Since the Google Play is not available in some countries or regions, these third-party app markets are the de facto official markets in these countries or regions.}, and
the Google Play.
In total, we have detected $967$ distinct FD apps and classified them into
$22$ families based on their code similarity. We then perform detailed analysis on the detected apps and observe some interesting findings, listed as following. 

\begin{itemize}
\item We found that most of the accounts in these apps
are chatbots, with fake user profile avatars. For instance, we find that the same user profile avatars are used by multiple different accounts in the same app, and even in different apps.
\item There are multiple parties involved in the ecosystem,
including app producers, app publishers, and distribution networks.
For example, we find that one developer key has been used to sign many FD apps with different package names, and
published by different companies with the same legal representative(s).
\item These apps are usually distributed through app markets and advertising networks. The fraudulent ranking techniques, e.g., fake user reviews and ratings
are used to manipulate the ranking of the apps (i.e., promote the apps).
\item We conduct an estimation of the overall revenue for the FD apps we have detected based on several reports.
Our estimation concludes that the total market scale is around 200 million US Dollars to 2 Billion US Dollars.
\end{itemize}

\noindent{\bf Contributions}\tab 
In summary, this paper makes the following main contributions:

\begin{itemize}
\item We have presented a new way adopted by malware authors to make a profit through luring users into buying
premium services in dating apps.

\item We have conducted a systematic study of FD apps and answered several key
research questions. To the best of our knowledge, our study is the first systematic
study of such kind of apps.

\item Our investigation has revealed various interesting findings that are previously unknown to the community.
We believe our study is the first step towards a better detection and regulation of such apps.
\end{itemize}

To engage the community, we will release all the FD apps we identified in this study and the experiment results to the research community for further analysis.

\section{Fraudulent Dating Apps Characterization}
\label{sec:characterizing}

 \begin{figure*}[t]
 \centering
 \includegraphics[width=6.5in]{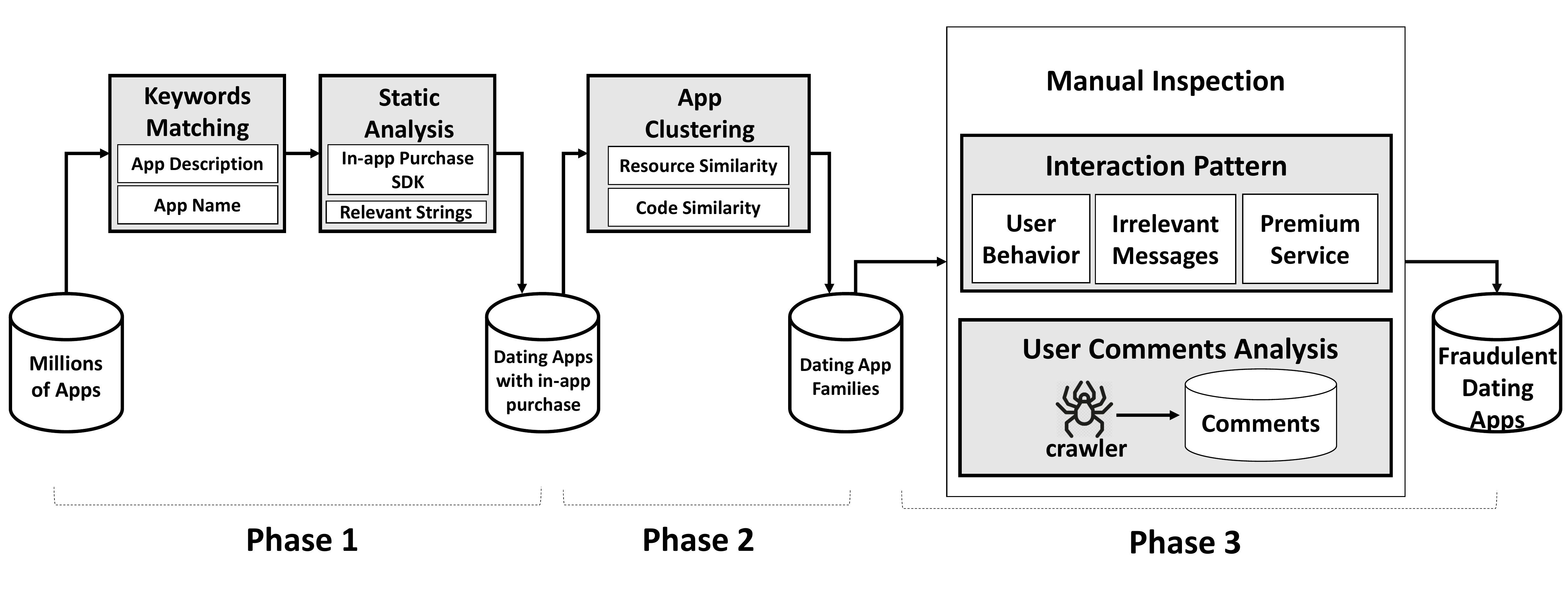}\\
  \caption{Our approach to identify FD apps.}
 \label{Figure:method}
 \end{figure*}

In this section, we first present our approach to identify FD apps in Section~\ref{subsec:app_identification}.
Then, in order to answer the following research question: \emph{Are the existing user accounts in these apps real persons or chatbots?},
we dissect the existing account profiles in different FD app families and then analyze
the interaction patterns between users of those dating apps in Section~\ref{subsec:user_profile} and Section~\ref{subsec:interaction}, respectively.

\subsection{Identifying Fraudulent Dating Apps}
\label{subsec:app_identification}
 
\subsubsection{Dataset} 

Table~\ref{table:dataset} presents an overview of our raw dataset that contains more than 2.5 million apps collected from ten Android app markets, including the official Google Play store. All of these apps were downloaded between April and August 2017.
We have also crawled the metadata of these apps, including app name, publisher company name, app version, rating, the number of downloads, etc.

\begin{table}[t]
\centering
\small
\caption{Overview of our dataset and the distribution of identified FD apps.}
\resizebox{0.8\linewidth}{!}{
\begin{tabular}{llcc}
\toprule
Market  & \#Apps & \# FD Apps & \# FD Apks\\ \midrule
Baidu Market & 227,454 & 673 & 3,227 \\
360 Market & 163,121 & 177 & 1,761 \\
Tencent Myapp & 636,265 & 432 & 2,782 \\
Xiaomi & 91,190 & 91 & 974 \\
Wandoujia & 554,138 & 285 & 2,767 \\
Huawei & 51,303 & 308 & 1,994 \\
Lenovo & 37,716 & 186 & 1,994  \\ 
OPPO & 426,419 & 305 & 1,628 \\
Meizu & 80,573 & 512 & 2,571 \\
Google Play & 287,110 & 7 & 123 \\
\midrule
Total & 2,555,199 & 967 & 3,697 \\
\bottomrule
\end{tabular}
}

\label{table:dataset}
\end{table}

\subsubsection{Methodology}

We propose a semi-automated approach to identify FD apps, as shown in Figure~\ref{Figure:method}.  
We first use a fast keywords matching method on the app metadata (e.g., app description and app name) to filter
dating app candidates from the millions of apps we have crawled. 
Then we perform static code analysis on the selected candidate apps to
check whether they have embedded in-app purchase services. 

The rationale behind this checking is that those services are essential for the app publisher to gain a profit, i.e., for victims to purchase premium services.
For the dating apps that embed in-app purchase services, we then cluster them based on resource similarity and code similarity. 
For apps in each cluster, we manually select several apps to inspect whether they have suspicious characteristics such as abnormal user behaviors and irrelevant messages, which make those apps as fraudulent app candidates.
We also analyze the user comments to further confirm that there exist victims of these apps in the real world.

\smallskip
\noindent{\bf Keywords Matching}\tab
Because the FD apps usually use seductive texts to attract victims,
we first collect several common words (in both English and Chinese) that frequently occur in the descriptions or app names of those apps,
including ``secret dating'', ``local single'', ``find girl'', etc. We use a fast keyword matching method to identify potential
dating app candidates.
Eventually, we are able to identify $61,133$ apps (out of 2.5 million apps) that contain at least two keywords.



\smallskip
\noindent{\bf In-app Purchase Analysis}\tab
One of the most important characteristics of FD apps is that they try to entice
users to purchase their premium services. 
For the selected dating app candidates, we perform static code analysis to detect whether they contain embedded in-app purchase services. If so, they will be identified as candidates for further analysis.

In our study,
we take advantage of LibRadar~\cite{libradar}, an open source obfuscation-resilient tool to identify
third-party libraries used in Android apps.  
We consider 18 popular third-party in-app purchases SDKs that are widely used in both China and worldwide, as shown in Table~\ref{table:paymentsdks}.

\begin{table}[t]
\centering
\small
\caption{Third-party in-app purchase services.}
\resizebox{1\linewidth}{!}{
\begin{tabular}{cc}
\toprule
Alipay & \url{https://www.alipay.com}\\
WeChatPay & \url{https://pay.weixin.qq.com/index.php/core/home/} \\
Paypal & \url{https://github.com/paypal/PayPal-Android-SDK}\\
YeePay & \url{https://www.yeepay.com}\\
Ping++ & \url{https://www.pingxx.com}\\
BaiduPay &\url{https://www.baifubao.com} \\
JieShenPay &\url{http://jieshenkj.com}\\
IPayNow & \url{Ipaynow.cn}\\
LianlianPay &\url{http://www.lianlianpay.com/international/} \\
UnionPay &\url{https://merchant.unionpay.com/join/index}\\
MengPay & \url{http://www.cnmengpay.com}\\
PayEco & \url{https://www.payeco.com}\\
SwiftPass & \url{http://www.swiftpass.cn}\\
JuHe & \url{https://www.ijuhepay.cn}\\
JuBaoPay & \url{http://www.jubaopay.com/\#/} \\
99Bill &\url{https://www.99bill.com} \\
IAppPay &\url{https://www.iapppay.com}\\
BBNPay &\url{https://www.bbnpay.com}\\
\bottomrule
\end{tabular}
}
\label{table:paymentsdks}
\end{table}

Note that, besides the third-party in-app purchase services,
the in-app purchase service provided by the Google Play is also considered in our study.
In some countries or regions where the Google Play service is not available,
app developers tend to use third-party in-app purchase services. 
For instance, the AliPay~\cite{alipay} and WeChatPay~\cite{wexpay} are the two most
popular third-party in-app payment services in China. 

Since some apps may implement their own payment functions (e.g., send SMS to premium number) instead of directly embedding third-party payment services,
we further investigate the related English and Chinese keywords (e.g., ``Purchase \& VIP'', ``Privilege'', etc.) in the layout configuration files to identify app candidates as supplementary. In total, we have identified $23,546$ dating apps that contain embedded in-app purchase services.

\begin{table}[t]
\newcommand{\tabincell}[2]{\begin{tabular}{@{}#1@{}}#2\end{tabular}}
\caption{An Overview of $22$ FD app families.}
\centering
\resizebox{1\linewidth}{!}{
\begin{tabular}{llc||llc}
\toprule
Family  & \# Apks & \#Pkg & Family  & \# Apks & \#Pkg \\ \midrule
Youyuan & 1,104  & 496  & Yuanlai & 70  & 12\\
Appforwhom  & 742 & 70 & Qianshoulian & 272& 15\\
Youairen  & 596   & 140  & Erwanshenghuo & 25 & 4\\
Yueaiapp  & 37 & 20  & Zlewx  & 16   & 5 \\
Tanliani   & 165 & 26 & Xiangyue  & 53  & 25 \\
Wmlover& 179  & 32  &99Paoyuan  & 44  & 5\\ 
Tongchengsupei & 50& 42 & Qiaiapp  & 27 & 13\\
Jiangaijiaoyou & 32&13&Meiguihunlian& 65  & 2\\
Aiaihunlian & 31  & 10  &Jucomic  & 22  & 8\\
Sipuhaiwei & 55  & 13  &Ailiaoba & 7 & 2\\
Yuanfenba & 51  & 11   &Michun   & 54  & 3\\
\midrule
Total &3,697  &967 \\
\bottomrule
\end{tabular}
}
\label{table:fraudfamily}
\end{table}

\noindent{\bf App Clustering}\tab
For the dating apps that embed in-app purchase services, we then cluster them based on resource similarity and code similarity. 

We first take advantage of the open source system FSquaDRA2~\cite{FSquaDRA2} to measure the
resource similarity of each app pair based on a feature set of resource names and asset signatures (the MD5 hash of each asset of an application excluding its icon and XML files).

We then use an app clone detection tool WuKong~\cite{wukong} to measure code-level similarity. 
For the apps with resource similarity scores higher than $90\%$ and code-level similarity scores higher than $85\%$\footnote{we choose the threshold empirically based on the previous studies}, we group them into the same cluster.
Finally, we select the cluster whose size is equal or bigger than 2. In total, we identified
$5,547$ candidate apps into $226$ clusters (size $>=2$).

\noindent{\bf Manual Inspection}\tab
For apps in each cluster, we manually select three apps\footnote{If the cluster size is 2, we select both apps.} (596 apps in total).
We installed them on smartphones and then registered real accounts to check whether they have the typical characteristics such as abnormal user behaviors
(many users will start conversations even though our registration information is empty or totally unattractive),
irrelevant messages and premium services. 

With the help of manual inspection, we eventually flag $22$ out of the $226$ clusters as FD app families.

\subsubsection{Statistics of Fraudulent Dating Apps}
As shown in Table~\ref{table:fraudfamily}, we have identified 3,697 FD apps (APKs)
that share 967 unique package names\footnote{One app (package name) corresponds to several APKs because our
crawler downloads different versions of apps during a 4 months span.}. These apps account for 6\% of dating apps in our dataset, which is much higher than we expected. 
For each family, we choose the keyword from the package name that has the most number of downloads as the family name.
For example, the family \code{Youyuan} includes the most number of unique package names and APKs, as roughly one
third of the APKs and more than half of the packages belong to this family.

The distribution of FD apps for different markets is shown in Table~\ref{table:dataset}. The Baidu Market hosts the most number of FD apps, where more than two-thirds of total FD apps are from this market. The official  Google Play market contains the least number of FD apps, where only 7 apps are flagged.

\subsection{User Profile Analysis}
\label{subsec:user_profile}


  \begin{figure}[t]
 \centering
 \includegraphics[width=3.2in]{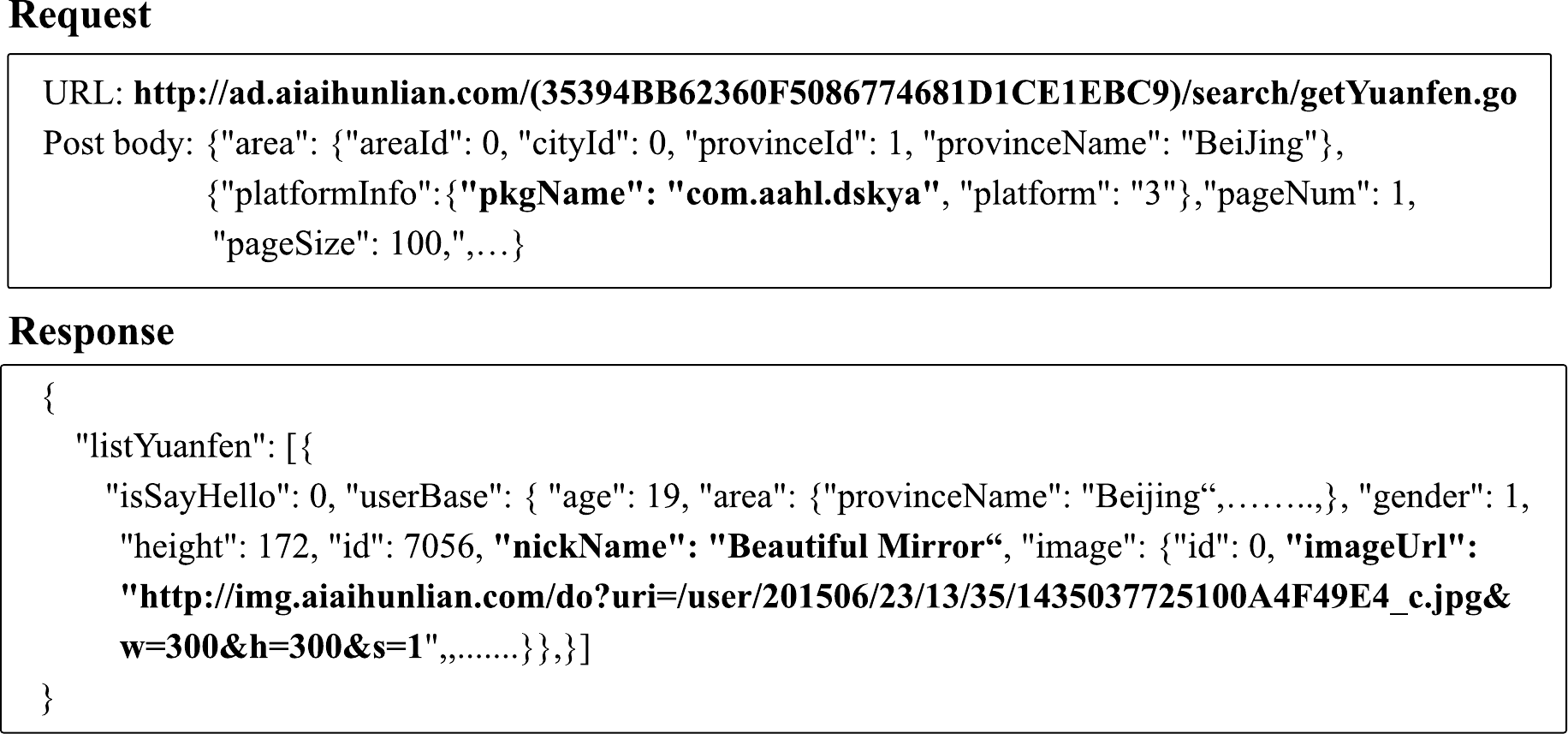}\\
 \caption{Protocol Analysis: the request and response messages to retrieve user information.}
 \label{Figure:protocol}
 \end{figure}

\subsubsection{Protocol Analysis}

It is non-trivial to harvest account profiles from FD apps, as they are not directly available on the devices.
We hence resort to the network traffic traces to retrieve user profiles.
Particularly, we randomly select three apps in each family and run them on real smartphones. 
We then leverage \emph{tcpdump} to record the
network traffic traces. Table~\ref{table:serveraddress} shows the server addresses of user profile requests and the download addresses of avatar files for each family. Surprisingly, all three apps we analyzed for each family share the same server address, even if they have different package names or developer signatures.

\label{appendix:address}
 \begin{table}[t]
\newcommand{\tabincell}[2]{\begin{tabular}{@{}#1@{}}#2\end{tabular}}
\caption{Server Address Analysis of 22 FD App Families.}
  \vspace{-0.1in}
\centering
\resizebox{1\linewidth}{!}{
\begin{tabular}{lll}
\toprule
Family & Addr of User List Request & Download Addr of Avatar\\ \midrule
Youyuan  & hulu.youyuan.com & ptw.youyuan.com \\
Appforwhom &aus.appforwhom.com &img0.appforwhom.com\\
Youairen& api.youairen.cn/api/users & cdnimg.365yf.com\\
Yueaiapp & napi.yueaiapp.cn& nimgup.yueaiapp.cn\\
Tanliani & api.tanliani.com& img.miliantech.com\\
Wmlover & app.wmlover.cn & cdn.wmlover.cn\\
Tongchengsupei & \tabincell{l}{jiaoyouappslb.tongcheng\\supei.cn} & \tabincell{l}{makefriends.oss-cn-qingdao.\\aliyuncs.com}\\
Jiangaijiaoyou & jiangaijiaoyou.com & image.jiangaijiaoyou.com\\
Aiaihunlian    & ad.aiaihunlian.com & img.aiaihunlian.com\\
Sipuhaiwei     & app.sipuhaiwei.com & piccdn.sipuhaiwei.com\\
Yuanfenba      & api2.app.yuanfenba.net & image.yuanfenba.net\\
Yuanlai        & mobileapi.yuanlai.com & photo4.ylstatic.com\\
Qianshoulian   & mpc5.qianshoulian.com & mpc5.qianshoulian.com\\
Erwanshenghuo  & api.erwanshenghuo.com & pic.erwanshenghuo.com\\
Zlewx          & zlewx.com & image.zlewx.com\\
Xiangyue       & v1.5xiangyue.cn & 7xkly7.com1.z0.glb.clouddn.com\\
99Paoyuan      & api2.99paoyuan.com & img7.kainei.com\\
Qiaiapp        & app.qiaiapp.com & photo.qiaiapp.com\\
Meiguihunlian  & api.meiguihunlian.com & photo.meiguihunlian.com\\
Jucomic        & yuemei.jucomic.com & photo.jucomic.com\\
Ailiaoba       & friend.ailiaoba.com.cn & liaobaimg1.mosheng.mobi\\
Michun         & api.michun.fallchat.com & \tabincell{l}{youyu-michun.oss-cn-shenzhen.\\aliyuncs.com}\\
\bottomrule
\end{tabular}
}
\label{table:serveraddress}
\end{table}

We further investigate the request and response messages for user information retrieval, as shown
in Figure~\ref{Figure:protocol}. The request messages usually contain information like geo-location data, platform information, the number of requested user information, etc.
The response messages contain a list of users where each of them is represented by a unique identifier, a URL of the avatar file, and some other personal information (e.g., nickname, age, etc.).

By deeply looking into those request and response messages, we observe that some apps (e.g., app \code{com.hzsj.qmrl} and app \code{com.wanjiang.tcyasq} in the app family \code{Youyuan}) embed a fingerprint or package name in the request message to differentiate the apps (Figure~\ref{Figure:protocol}).
Moreover, some apps (e.g., app \code{com.yuanfenapp.tcyyjiaoyou} and app \code{ltd.onedream.snsapp.moaiyueai} in the app family \code{Tongchengsupei}) may share exactly the same request and response messages, resulting in identical accounts. 

\subsubsection{Crawling Account Profiles}

To crawl the account profiles, one straightforward approach is to simulate the protocol for each app.
However, because account information is usually shown based on geo-location (e.g., you could only browse the user locations in the same city), and each request could only get limited number of users (e.g., one page), we need to analyze the request URLs to identify the corresponding fields, and construct request messages (e.g., change the city or the page number of user profiles) so as to crawl as many as possible user profiles. 

Unfortunately, app developers could use anti-crawling techniques such as embedding hash values in the request URLs,
to keep us away from automatically harvesting their account profiles. 
Due to this reason, we propose to employ automated app testing techniques to infer user profiles. 
In particular, we leverage an automated UI testing tool DroidBot~\cite{droidbot} to generate UI pull-down events and send to the tested apps to emulate real user behaviors of browsing the account list.

 \begin{figure*}[t]
 \centering
 \includegraphics[width=6.5in]{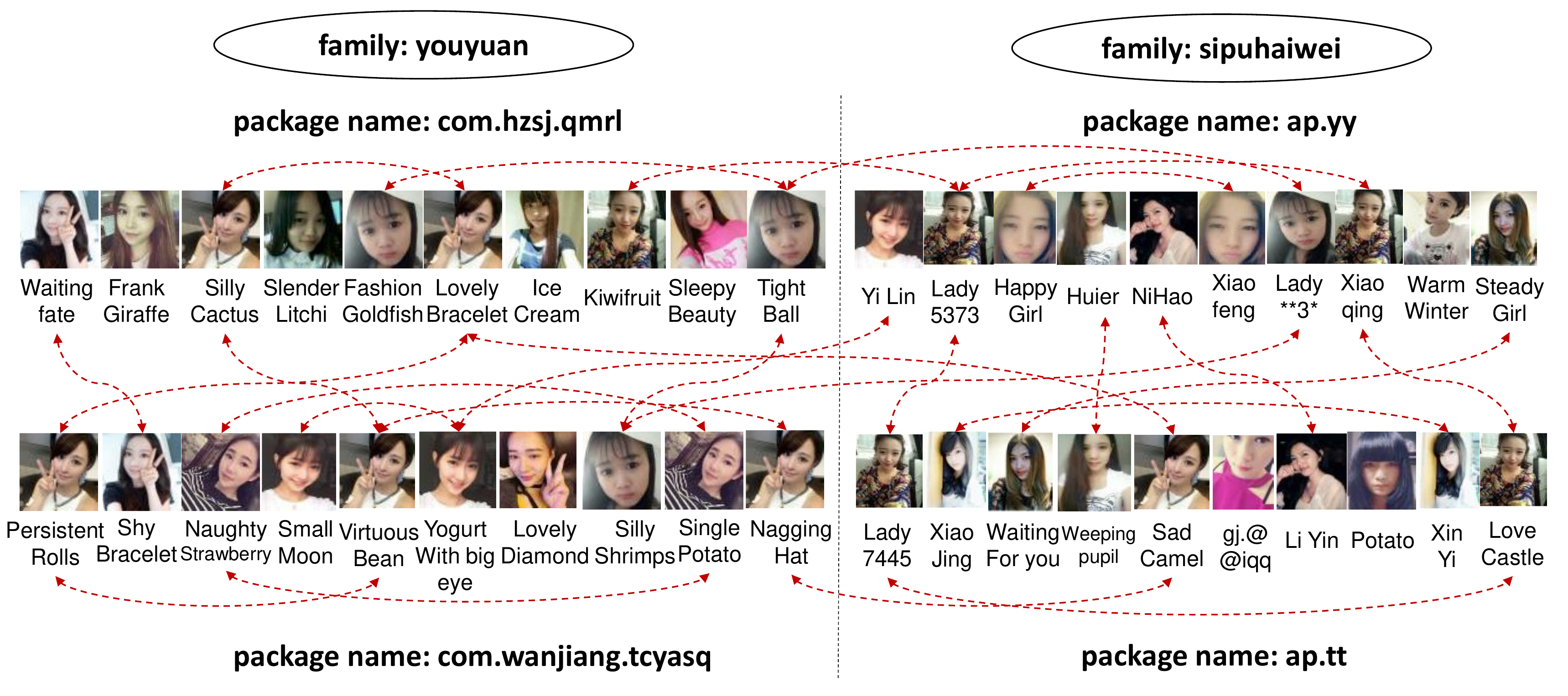}\\
 \caption{Examples of fake account profiles.}
 \label{Figure:fakeProfile}
 \end{figure*}
 
\subsubsection{The Presence of Fake Account Profiles}

It is difficult to measure how many fake accounts actually exist in each app since it is impossible for us to start
a conversation with each account to check whether he/she is a real person or not. 
Motivated by the characteristics of Romance Scam fraud~\cite{romancescam} that the scammers usually
post profiles using stolen photographs of attractive people, we believe the fake accounts in FD apps
may also use stolen/online photos too. To identify fake accounts in a fast manner,
we regard the accounts with the same avatar photos but totally different account
information (e.g., nickname, hometown, age, etc.) within the same app  as \emph{fake accounts}.

In this study, we use Dup Detector~\cite{dupdetector}, a pixel-level comparison technique to identify duplicate
images. Note that some apps offer default avatar photos during registration, which could mislead our detection. Thus we exclude all the default avatars from photo comparison.
For each family, we randomly choose an app and crawl all the account profiles. As shown in Table~\ref{table:fakeprofilepackage},
for the app \code{com.jqyuehui.main} belonging to the \code{Youairen} family, we are able to crawl over 263,000 account profiles.

\begin{table}[t]
\newcommand{\tabincell}[2]{\begin{tabular}{@{}#1@{}}#2\end{tabular}}
\caption{The distribution of fake account profiles within the same app for each family.}
\centering
\resizebox{1\linewidth}{!}{
\begin{tabular}{lccc}
\toprule
Package Name & \# Account Profiles & \# Fake Account Profiles & Percent \\ \midrule
Youyuan  & 57,779 & 7403 & 12.81\% \\
Appforwhom & 5,108 & 130 & 2.55\% \\
Youairen& 263,932 & 153,842 & 58.29\% \\
Yueaiapp & 2245 & 6 & 0.27\%\\
Tanliani & 969 & 0 & 0.00\% \\
Wmlover & 5516 & 683 & 12.38\% \\
Tongchengsupei & 3663 & 42 & 1.15\% \\
Jiangaijiaoyou & 1872 & 0 & 0.00\% \\
Aiaihunlian    & 663 & 4 & 0.60\% \\
Sipuhaiwei     & 23,946 & 1646 & 6.87\% \\
Yuanfenba      & 3864 & 25 & 0.65\% \\
Yuanlai        & 2178 & 8 & 0.37\% \\
Qianshoulian   & 5431 & 312 & 5.74\% \\
Erwanshenghuo  & 474 & 0 & 0.00\%\\
Zlewx          & 3410 & 30 & 0.88\% \\
Xiangyue       & 6451 & 38 & 0.59\% \\
99Paoyuan      & 7829 & 411 & 5.25\%\\
Qiaiapp        & 3422 & 2 & 0.06\%\\
Meiguihunlian  & 7132 & 127 & 1.78\%\\
Jucomic        & 4173 & 43 & 1.03\% \\
Ailiaoba       & 3348 & 2 & 0.06\% \\
Michun         & 5948 & 0 & 0.00\%\\
\bottomrule
\end{tabular}
}
\label{table:fakeprofilepackage}
\end{table}

Figure~\ref{Figure:fakeProfile} shows examples of fake account profiles we have found in our crawled data, from which we can observe that fake accounts may exist within the same app, within the same family, or even across different families.

\smallskip
\noindent{\bf Fake Accounts within the Same App}\tab
We first measure the fake account profiles within each app. As shown in Table~\ref{table:fakeprofilepackage}, although we have identified fake account profiles in most of the apps, the percentage of fake account profiles within each app is not high. Only three apps have more than 10\% of their account profiles detected as fake ones. Most of the apps have less than 1\% of fake account profiles. 

Note that, this is a
conservative way to identify fake account since different fake accounts inside one app
could use different avatar photos. We will
 analyze the interaction patterns in Section~\ref{subsec:interaction} to further detect the fake accounts.

\smallskip
\noindent{\bf Fake Accounts within the Same Family}\tab
For each family, we choose three apps to examine fake account profiles across apps but within the same family.

Since it is generally time-consuming to analyze the protocol of a given app to simulate the request messages, in this work, we select four popular families (12 apps in total) to perform our measurements. For every app family considered, we ensure that the apps inside the family are different (i.e., has different package names). 
The type of account information (e.g., age, location) may slightly vary across different apps, we therefore regard the accounts with same avatar photos but different nicknames as \emph{suspicious fake accounts.}

As shown in Table~\ref{table:fakeprofilefamily}, the rate of suspicious fake accounts within the same family is
significantly high. For example, around 95\% of account photos of the selected three apps in \code{Youyuan} family are overlapped.
The ratio of overlapping account profiles in \code{Appforwhom} family even achieves 100\%.
Further analysis reveals that all the apps in this family use
the same protocol for accessing account information.

\begin{table}[t]
\newcommand{\tabincell}[2]{\begin{tabular}{@{}#1@{}}#2\end{tabular}}
\caption{The distribution of fake account profiles within the same family.}
\centering
\resizebox{1\linewidth}{!}{
\begin{tabular}{lccc}
\toprule
Family & \# Account Profiles & \# Fake Account Profiles & Percent \\ \midrule
Youyuan  & 137,497 & 130,558 & 94.95\% \\
Appforwhom & 15,324 & 15,324 & 100\% \\
Youairen& 431,697 & 237,476 & 55.01\% \\
Sipuhaiwei & 58,194 & 36,642 & 62.97\%\\
\bottomrule
\end{tabular}
}
\label{table:fakeprofilefamily}
\end{table}

\smallskip
\noindent{\bf Fake Accounts Across Families}\tab
We measure the ratios of account profiles overlapping across different families. As shown in Table~\ref{table:fakeprofilefamily}, the overlapping ratios across different families are not high, where all of them are less than 10\%. 
One possible reason is that apps in different families are from different developers and have different sources of account profiles. 

\begin{table}[t]
\newcommand{\tabincell}[2]{\begin{tabular}{@{}#1@{}}#2\end{tabular}}
\caption{The distribution of fake account profiles across families.}
\centering
\resizebox{1\linewidth}{!}{
\begin{tabular}{lcccc}
\toprule
Family & Youyuan & Appforwhom & Youairen & Sipuhaiwei \\ \midrule
Youyuan  & 7403(12.81\%) & 137(2.68\%) & 3186(1.21\%) & 1435(5.99\%) \\
Appforwhom & 133(0.23\%) & 130(2.56\%) & 712(0.27\%) & 56(0.23\%)\\
Youairen& 3305(5.72\%) & 751(14.7\%) & 153842(58.29\%) & 2152(8.99\%) \\
Sipuhaiwei & 1316(2.28\%) & 58(1.14\%) & 1897(0.72\%) & 1646(6.87\%)\\
\bottomrule
\end{tabular}
}

\label{table:fakeprofileoverlap}
\end{table}

\begin{figure*}[t]
 \centering
 \includegraphics[width=6.2in]{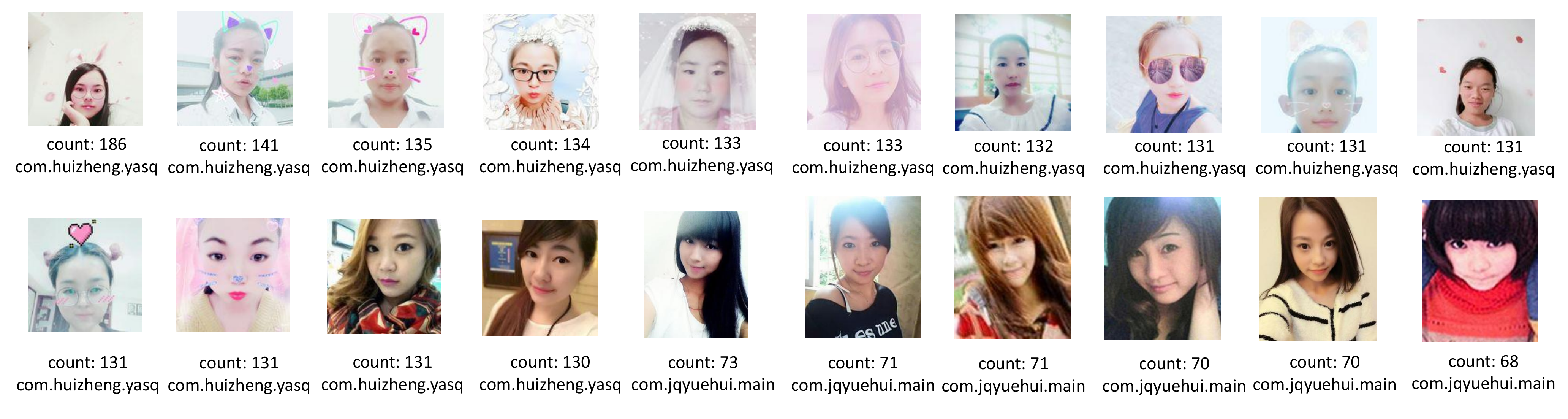}\\
 \caption{Top 20 mostly used fake user avatars.}
 \label{Figure:popularphoto}
 \end{figure*}

\smallskip
\noindent{\bf Mostly Used Fake User Avatars}\tab
We found many fake accounts using the same avatar photos but totally different account information. We list the top 20 popular fake user photos in Figure~\ref{Figure:popularphoto}. All of the top 14 apps belong to \code{com.huizheng.yasq} (family \code{Youyuan}), while all of them have appeared in at least 130 different accounts.
We also calculate the number of different accounts with each avatar photo and show it under the avatar in the Figure.

\begin{table*}[t]
\newcommand{\tabincell}[2]{\begin{tabular}{@{}#1@{}}#2\end{tabular}}
\caption{Field study of different FD app families.}

\centering
\resizebox{\linewidth}{!}{
\begin{tabular}{lccccccc}
\toprule
Family (package name) & \tabincell{c}{Phone Num\\SN account\\Email} &\tabincell{c}{Template\\based}& \# \tabincell{c}{Chatups\\(M)} & \# \tabincell{c}{Chatups\\(F)} & \# \tabincell{c}{Free\\Messages} & \tabincell{c}{Content\\Relevance} & \tabincell{c}{After\\Purchase}\\ \midrule
Youyuan (com.huizheng.dsya)  & \XSolid & \Checkmark & 6 & 0 & \XSolid & NA & \XSolid \\
Appforwhom (cn.com.qncnew.aus)  & \XSolid & \XSolid  & 7 &  5 & \XSolid & NA & \XSolid\\
Youairen (com.jqyuehui.main) & \XSolid & \XSolid  & 3 & 0 & Total 3 & \XSolid & \XSolid \\
Yueaiapp (com.yueai.ya007)  & \XSolid & \XSolid  & 12 & 0 & 1 Per User & \XSolid & \XSolid\\
Tanliani (com.blsm.miyou)  & \XSolid & \XSolid  & 6 & 0 & Total 3 & NA & \XSolid\\
Wmlover (cn.umuad.dsaq)  & \XSolid & \XSolid  & 5 &0 & \XSolid & NA & \XSolid\\
\tabincell{l}{Tongchengsupei\\(ltd.onedream.snsapp.moaiyueai)}  & \XSolid &  \XSolid  & 5 &0 & \XSolid & NA & \XSolid\\
\tabincell{l}{Jiangaijiaoyou \\(com.jiangaihunlian.danshenyuehui)}  & \XSolid & \XSolid  & 6 & 0& \XSolid & NA & \XSolid\\
Aiaihunlian (com.aahl.jmyhb)     & \XSolid & \Checkmark & 3 &0 & \XSolid & NA & \XSolid \\
Sipuhaiwei (com.myhoney)  & \XSolid & \XSolid  & 14 &0 & \XSolid &NA & \XSolid\\
Yuanfenba (com.xiaochen.android.fate\_it)      & \XSolid & \XSolid  & 4 &0 & \XSolid & NA & \XSolid\\
Yuanlai (com.yuanlai) & \Checkmark  & \XSolid  & 7 &0 & \XSolid & NA & \XSolid \\
Qianshoulian (com.xiangqinqin.app)& \XSolid & \XSolid  & 2 &0 & 1 Per User & \XSolid & \XSolid\\
Erwanshenghuo (com.syty.todayDating) & \XSolid & \XSolid  & 3 &0 & \XSolid & NA &\XSolid \\
Zlewx (com.jshy.tongcheng)  & \XSolid & \XSolid  & 5 &0 & \XSolid &NA &\XSolid \\
Xiangyue (com.luren.xiangyueai) & \XSolid & \Checkmark & 5 &0 & \XSolid &NA & \XSolid \\
99Paoyuan (com.lingnei.kaikai) & \Checkmark & \XSolid  & 6 &\XSolid & \XSolid &NA & \XSolid \\
Qiaiapp (com.jiaoyou.jqya)& \XSolid & \XSolid  & 4 &0  & \XSolid &NA & \XSolid\\
Meiguihunlian (com.meiguihunlian) & \XSolid & \XSolid  & 4 &0 & \XSolid &NA & \XSolid\\
Jucomic (cn.nineox.yuejian)& \Checkmark & \XSolid  & 2 &0 & \XSolid &NA &  \XSolid\\
Ailiaoba (com.liaoba)& \Checkmark & \XSolid  & 3 &\XSolid & 1 Per User &NA &\XSolid \\
Michun (com.youyu.michun)& \XSolid & \XSolid  & 8 &0 & 1 Per User &\XSolid &\XSolid \\
\bottomrule
\end{tabular}
}
\label{table:Interaction}
\end{table*}

\subsection{Interaction Pattern Analysis}
\label{subsec:interaction}

We now attempt to identify fake accounts from another perspective, i.e., the interaction patterns.
If the accounts are real persons, then the messages should be relevant to the topic of the conversation.
Thus, we perform a field study to analyze the interaction patterns of these fraudulent dating apps.
For each family, we randomly choose an app and install it on a real device.
Then we register two accounts (1 male user and 1 female user) to log in and start a conversation.
Furthermore, we purchase the premium service  for each app and compare the results
before and after purchasing their services.

As shown in Table~\ref{table:Interaction}, we have observed several interesting findings:
 \begin{enumerate}
 \item \emph{The registration process is quite easy, and most apps do not need any personal information}. As shown in the second column of Table~\ref{table:Interaction}, only 4 (out of 22) apps require the phone number, social networking or email account during registration. 

\item \emph{Several apps use template-based conversations.} As shown in the third column of Table~\ref{table:Interaction}, 3 (out of 22) apps use template-based conversations, which could be found in the resource files of the app. 

\item \emph{There is a huge difference between male users and female users}. For male users, when they are online, many female accounts will reach to them within a short time (see column \#4). For example, more than 10 girls talked to our registered user for the app \code{com.yueai.ya007} (family: \code{Yueaiapp}) and
 \code{com.myhoney} (family: \code{Sipuhaiwei}) within five minutes during our experiment. 
However, for female users, there was no one trying to initiate conversation for almost all the apps during our experiment (see column \#5).
Specifically, for some apps (e.g., \code{com.liaoba}), the 
the default gender is male during registration and users cannot change it.
This suggests that these apps are mainly targeting male users.

 \item \emph{For more than 70\% of the apps, users cannot reply to the messages unless premium services are purchased}.
 For the remaining 6 apps, users could only respond to at most 3 messages. 
 
 \item \emph{Irrelevant messages are prevalent in the conversations.} Before we purchased the premium services, we were only able to reply to the messages in 6 apps. However, only 4 accounts in these apps replied to us and the response messages were  totally irrelevant.
 
 \item \emph{After purchasing the premium services, the app stops responding to messages.} In this field study, we spent roughly 176 US dollars to purchase premium
 services for these 22 apps. Unfortunately, once we had purchased the premium services, all apps stopped responding to our messages. It appears that the sole mission of these apps is to lure users into purchasing its so-called premium services, which in reality do not exist at all. 
 \end{enumerate}

\subsection{Summary}

Based on the results of user profile and interaction pattern analysis, we suspect that the accounts (except for the victims)
in the apps are chatbots, instead of real persons. First, the account profiles in different apps in a family are mostly identical.
These account profiles may be automatically generated. Second, our suspicion can be further confirmed by the
interaction patterns. For instance, the messages in the conversation for the apps we evaluated are irrelevant to the topic, and no messages
will be received after purchasing the premium services. These patterns are more likely generated from computer programs instead of real persons.

\section{Business Model Analysis}
\label{sec:business}

We now analyze the business model of FD apps aiming at
revealing the involved parties in the ecosystem, so as to answer the research question: \emph{what are the involved parties and how they make a profit?}

We first retrieve
the signature of the developer's key inside the app. If the signatures are the same in two different apps, 
we assume that these two apps are developed by the same developer\footnote{This is a reasonable assumption since the leakage of the developer's key is not a common case in practice.}.
These developers are referred as \code{app producers}.

We then collect the released company names of the FD apps from the
app markets. Usually, the company name is a required entry when publishing apps to an app market. 
We also collect the name of the legal representative~\cite{legalperson} of the company,
based on the public records from the corresponding government agencies.
These companies are the ones who 
publish apps and obtain payments from victims.
We call them \code{app publishers}.

\begin{table}[t]
\newcommand{\tabincell}[2]{\begin{tabular}{@{}#1@{}}#2\end{tabular}}
\caption{Multiple parties in the FD app ecosystem.}
\centering
\resizebox{1\linewidth}{!}{
\begin{tabular}{l|c|c|cc}
\toprule
Family & \#Pkg &\# \tabincell{c}{Developer\\ Signature} & \# Companies & \#\tabincell{c}{Legal\\Persons} \\ \midrule
Youyuan  & 496& 48   & 113   & 107  \\
Appforwhom &70& 43  & 21  & 10 \\
Youairen & 140& 133  &  27 & 21\\
Yueaiapp & 20 &5  &  4 & 4\\
Tanliani & 26& 10  & 11  & 10\\
Wmlover & 32 & 31 &  9 & 9 \\
Tongchengsupei &42 & 5  & 16  & 16 \\
Jiangaijiaoyou &13& 6  & 7  & 7\\
Aiaihunlian    &10& 5  & 3  & 3 \\
Sipuhaiwei     &13& 8  & 5  & 5\\
Yuanfenba      &11& 10  & 5  & 5\\
Yuanlai        &12&  1 & 3  & 3\\
Qianshoulian   &272&  12 & 7  & 7\\
Erwanshenghuo  &25&  1 & 3  & 3\\
Zlewx          &16&  1 & 2  & 2\\
Xiangyue       &53&  22 & 6  & 5 \\
99Paoyuan      &44&  5 & 3  & 3 \\
Qiaiapp        &27&  3 & 6  & 6\\
Meiguihunlian  &65&  2 & 1  & 1\\
Jucomic        &22&  6 & 3  & 3\\
Ailiaoba       &7&  1 & 2 & 2\\
Michun         &54& 3  & 1  & 1\\
\bottomrule
\end{tabular}
}
\label{table:multiparty}
\end{table}

Table~\ref{table:multiparty} shows the data of multiple parties involved in the
FD app ecosystem. In particular, the first column shows the family
name of the app, and the second column shows the number of distinct package names
in each family. The third column shows the number of distinct developer signatures for the
apps in each family, while the last two columns show the number of distinct
company names and that of legal representatives of the companies. 
The number of legal representatives is less or equal to the number of companies since the same person can serve as the legal representatives of multiple companies.

Based on the data in Table~\ref{table:multiparty}, we obtain the following
observations. 
\begin{itemize}

\item The number of developer signatures is usually much fewer
than the number of distinct package names in each family.
This evidence indicates that the developers of different apps may be the same person. 
We also find the case that even the developer signatures of some apps are different, the names of the RSA files inside the META-INF directory of those apps are identical.

For example, there are $130$ apps in the family \code{youairen} sharing the same signature file name, namely \emph{KEY\_KEYS.RSA}.
For each signature, the values of the \code{CN} (Common Name) and \code{OU} (Organization Unit) fields are meaningless strings such as \code{estituan}, \code{umgfoubq}.
We believe even though the signatures of these $130$ keys are different, they are probably automatically generated by the same person.
Table~\ref{table:key} shows the randomly generated \code{CN} and \code{OU} fields of $20$ developers' signatures.

\item  The apps belonging to the same family are published by multiple companies.
This could be explained by the fact that the producers may sell their apps to different companies for publishing. For instance,
as shown in the code snippet of Listing~\ref{listing:merchantID}, the app developer hard-coded the relationship between the package names
and the payment accounts that are used to receive payments from victims.
By doing so, app producers could sell the same app with different package names to different companies (or publishers).

\item One legal representative could own multiple companies so that FD apps can be published multiple times via different companies.
If the apps from one company are removed from app markets due to user complains, the apps from
other companies can still survive.
\end{itemize}


\begin{table}[t]
\newcommand{\tabincell}[2]{\begin{tabular}{@{}#1@{}}#2\end{tabular}}
\caption{The randomly generated OU and CN field in the developers' signatures.}
\centering
\resizebox{1\linewidth}{!}{
\begin{tabular}{c|c||c|c||c|c||c|c}
\toprule
OU            &       CN        &  OU &         CN  &OU            &       CN    &OU            &       CN     \\ \midrule
hyuhzwhl & fstxzj & btxetuui &  pfsmdm  & oymorwlp & vrnnuv & estituan & hvhskr \\
umgfoubq & vvshla & mdjgjpuc &  tkwcpk  & agggefmk & zhmipw & kovkokxi & jzmmeh \\
pbvdysyg & kychif & uzqoaawn &  usfzum  & gfpsbhmz & sejwql & dgvcmhxh & sqsqrf \\
pebbomoy & dacyfr & xwvoirgr &  pgupio  & vdsqjkvk & alzrly & ugqljbld & swkcom  \\
memqwnon & nmscvv & wcjhpfkb &  lxkruk  & xdttilqo & zaamyd & hrzwtnzj & hvsbwp \\

\bottomrule
\end{tabular}
}
 \vspace{-0.1in}
\label{table:key}
\end{table}

\begin{minipage}[t]{0.45\textwidth}
  \centering
\begin{lstlisting}[language=Java, caption=The code snippet showing the app package names and their corresponding WeChatPay accounts used to receive payments from victims, label={listing:merchantID}]
str=package_name;
if ("com.huizheng.lasq".equals(str)) {
   com.app.tencent.QQConstants.APP_ID="wxba66413d0f792ffa";
   com.app.tencent.QQConstants.PARTNER_ID="1296280201";
   return;
}
if ("com.youyuan.lrxq".equals(str)) {
   com.app.tencent.QQConstants.APP_ID="wxf54f037e28294cc6";
   com.app.tencent.QQConstants.PARTNER_ID="1296283001";
   return;
}
...
\end{lstlisting}
\end{minipage}

 \begin{figure*}[t]
 \centering
 \includegraphics[width=5in]{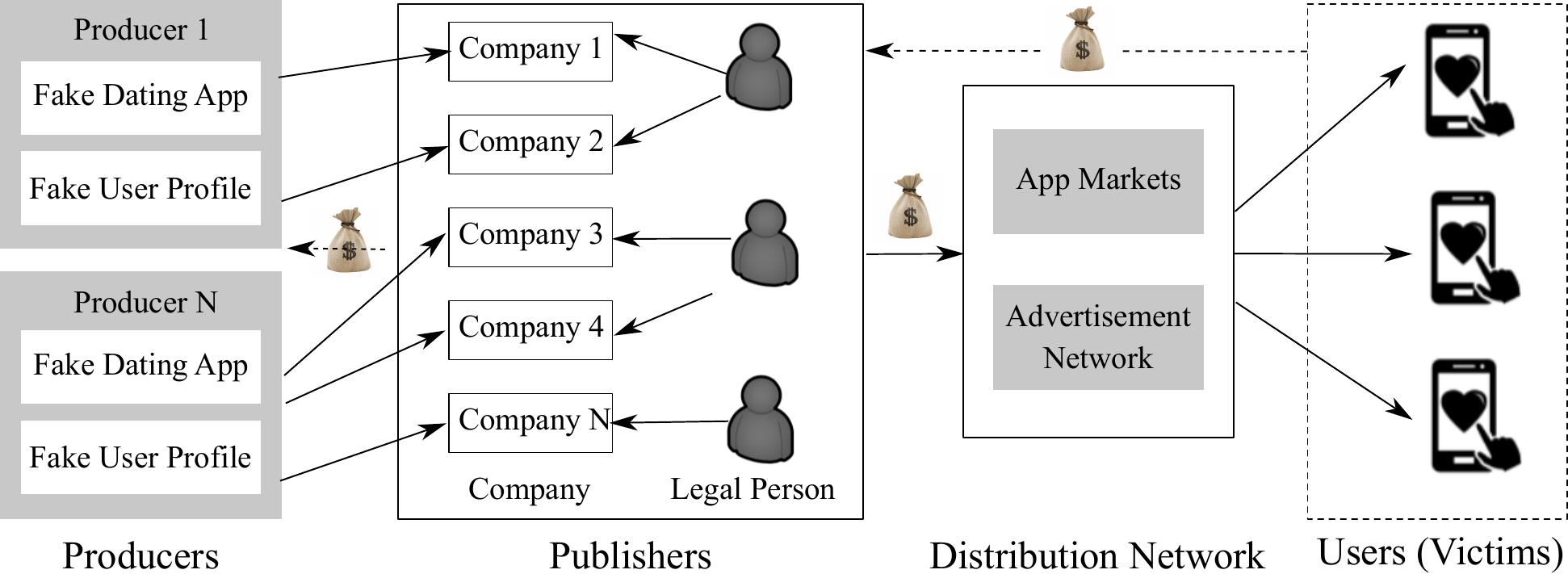}\\
 \caption{The business model of the FD apps.}
 \label{Figure:business}
 \end{figure*}

Figure~\ref{Figure:business} shows the business model of the ecosystem.
Specifically, the app producers develop apps and sell them to publishers. The publishers usually register multiple companies and use these companies to distribute their apps, e.g., via app markets. In order
to promote these apps, the ranking fraud techniques are used (Table~\ref{table:rankingfraud} in Section~\ref{subsec:appmarket}).
Moreover, the publishers could also pay the advertising network to distribute their apps (Section~\ref{subsec:adnetwork}). 
When the victims are lured into installing these apps and buying the premium services,
publishers will receive the money and gain a profit.
Note that, the producers and publishers in some cases may come from the same companies,
and act as both roles in the ecosystem.

\section{Distribution Network Analysis}
\label{sec:distribution}

\begin{table*}[t]
\newcommand{\tabincell}[2]{\begin{tabular}{@{}#1@{}}#2\end{tabular}}
\caption{Ranking fraud based on user reviews.}
\centering
\resizebox{\linewidth}{!}{
\begin{tabular}{l|cc|cccc|llc}
\toprule
 
Family & \# \tabincell{c}{Reviews} & \# \tabincell{c}{Reviews \\with Five-star \\Rating}  & \#  \tabincell{c}{Repeated\\Reviews} & \# \tabincell{c}{Fake\\Reviews} &\% \tabincell{c}{Fake\\Reviews} & \% \tabincell{c}{Fake Reviews\\with Five-star\\Rating}  & \# Users & \# \tabincell{c}{Fake\\Users}&\% \tabincell{c}{Fake\\Users}\\ \midrule
Youyuan  &        \fnum{354797  }  &  \fnum{337140 } & \fnum{188811   }   & \fnum{285448 }    & {80.45\% } &  {95.82\%}   & \fnum{195256 } & \fnum{145151 } & {74.34\% }\\
Appforwhom &      \fnum{94693   }  &  \fnum{91760  } & \fnum{78382    }   & \fnum{89945  }    & {94.99\% } &  {98.73\%}   & \fnum{26603  } & \fnum{23799  } & {89.46\% }\\
Youairen &        \fnum{380722  }  &  \fnum{379560 } & \fnum{221936   }   & \fnum{347486 }    & {93.76\% } &  {99.26\%}   & \fnum{135710 } & \fnum{121801 } & {89.75\% }\\
Yueaiapp &        \fnum{20920   }  &  \fnum{19562  } & \fnum{13490    }   & \fnum{17990  }    & {85.99\% } &  {98.63\%}   & \fnum{15895  } & \fnum{13811  } & {86.89\% }\\
Tanliani &        \fnum{51951   }  &  \fnum{52032  } & \fnum{31427    }   & \fnum{44812  }    & {86.26\% } &  {97.74\%}   & \fnum{38154  } & \fnum{29835  } & {78.20\% }\\
Wmlover &         \fnum{103162  }  &  \fnum{100183 } & \fnum{71787    }   & \fnum{91081  }    & {88.29\% } &  {99.25\%}   & \fnum{62352  } & \fnum{52005  } & {83.41\% }\\
Tongchengsupei &  \fnum{4746    }  &  \fnum{4594   } & \fnum{2007     }   & \fnum{4230   }    & {89.13\% } &  {99.21\%}   & \fnum{3959   } & \fnum{3479   } & {87.88\% }\\
Jiangaijiaoyou &  \fnum{3043    }  &  \fnum{1733   } & \fnum{890      }   & \fnum{1397   }    & {45.91\% } &  {78.38\%}   & \fnum{2195   } & \fnum{1002   } & {45.65\% }\\
Aiaihunlian    &  \fnum{19299   }  &  \fnum{18864  } & \fnum{11124    }   & \fnum{18406  }    & {95.37\% } &  {99.09\%}   & \fnum{12179  } & \fnum{11439  } & {93.92\% }\\
Sipuhaiwei     &  \fnum{46036   }  &  \fnum{45337  } & \fnum{27691    }   & \fnum{44292  }    & {96.21\% } &  {99.56\%}   & \fnum{15603  } & \fnum{14287  } & {91.57\% }\\
Yuanfenba      &  \fnum{21065   }  &  \fnum{20426  } & \fnum{12057    }   & \fnum{19899  }    & {94.46\% } &  {98.73\%}   & \fnum{15747  } & \fnum{14814  } & {94.08\% }\\
Yuanlai        &  \fnum{18702   }  &  \fnum{17126  } & \fnum{14991    }   & \fnum{16221  }    & {86.73\% } &  {97.95\%}   & \fnum{8429   } & \fnum{6558   } & {77.80\% }\\
Qianshoulian   &  \fnum{38514   }  &  \fnum{37006  } & \fnum{24571    }   & \fnum{34867  }    & {90.53\% } &  {98.91\%}   & \fnum{26346  } & \fnum{23509  } & {89.23\% }\\
Erwanshenghuo  &  \fnum{174     }  &  \fnum{119    } & \fnum{114      }   & \fnum{114    }    & {65.52\% } &  {85.96\%}   & \fnum{162    } & \fnum{118    } & {72.84\% }\\
Zlewx          &  \fnum{1004    }  &  \fnum{959    } & \fnum{775      }   & \fnum{921    }    & {91.73\% } &  {99.11\%}   & \fnum{972    } & \fnum{911    } & {93.72\% }\\
Xiangyue       &  \fnum{13408   }  &  \fnum{13108  } & \fnum{10604    }   & \fnum{12865  }    & {95.95\% } &  {99.36\%}   & \fnum{9645   } & \fnum{9257   } & {95.98\% }\\
99Paoyuan      &  \fnum{9505    }  &  \fnum{8954   } & \fnum{6126     }   & \fnum{8529   }    & {89.73\% } &  {97.74\%}   & \fnum{7428   } & \fnum{6652   } & {89.55\% }\\
Qiaiapp        &  \fnum{8955    }  &  \fnum{8521   } & \fnum{6537     }   & \fnum{7362   }    & {85.23\% } &  {98.51\%}   & \fnum{5780   } & \fnum{5088   } & {88.03\% }\\
Meiguihunlian  &  \fnum{1692    }  &  \fnum{1506   } & \fnum{758      }   & \fnum{1289   }    & {76.18\% } &  {91.81\%}   & \fnum{1237   } & \fnum{1013   } & {81.89\% }\\
Jucomic        &  \fnum{1540    }  &  \fnum{1428   } & \fnum{836      }   & \fnum{1134   }    & {73.64\% } &  {98.21\%}   & \fnum{1285   } & \fnum{996    } & {77.51\% }\\
Ailiaoba       &  \fnum{626     }  &  \fnum{592    } & \fnum{66       }   & \fnum{306    }    & {48.88\% } &  {97.45\%}   & \fnum{611    } & \fnum{300    } & {49.10\% }\\
Michun         &  \fnum{6239    }  &  \fnum{5915   } & \fnum{4947     }   & \fnum{5759   }    & {92.31\% } &  {98.15\%}   & \fnum{2292   } & \fnum{2143   } & {93.50\% }\\
\midrule
Total          & \fnum{1201432}    &  \fnum{1166425  } &\fnum{729927  }   & \fnum{1054623} & {83.97\%} & {96.70\%}        & \fnum{483231} & \fnum{383181} & {82.92\%} \\
\bottomrule
\end{tabular}
}
\label{table:rankingfraud}
\end{table*}

App publishers usually distribute their apps
through both app markets and advertising networks (Figure~\ref{Figure:business}). In this section, we analyze
the distribution of the FD apps and answer the following question: \emph{how these apps
are distributed, and what techniques are used by publishers to promote
these apps in app markets?} 
In particular, we collect the names and user reviews of
these apps in app markets and monitor the app distribution through an online service~\cite{appgrowing}. 


\vspace{-0.1in}
\subsection{App Markets}
\label{subsec:appmarket}
As expected, we find that app markets are the primary choice for app publishers to
distribute apps (see Table~\ref{table:dataset} for the app markets
in which FD apps are detected).
Unfortunately, the ranking system of app markets could be manipulated (known as ranking fraud) by the owners of FD apps so as to attract more victims.

\smallskip
\noindent{\bf Ranking Fraud}\tab
Ranking fraud~\cite{rankingfraud} refers to the behaviors that aim to promote the ranking of apps inside
app markets. Based on the ranking mechanisms of app markets, this could be achieved
by manipulating the user reviews and rating of an app.
For each app in our study, we therefore crawl the reviews and the ratings (if it is available) of the app in each market as well as the names of users who have posted the reviews, aiming at identifying fake reviews and fake reviewers.

Specifically, our analysis is based on the following reasonable heuristic: reviews from different users
should be different in most cases. Though some simple reviews such as \emph{``great app''}
could be posted by different users, other reviews with more meaningful words
should not be exactly the same. Based on this heuristic, our analysis works in the following steps and the overall result is shown in Table~\ref{table:rankingfraud}.

First, we remove the reviews that have less than 5 words from our analysis to avoid potential false positives introduced by simple reviews.
The number of reviews, and reviews with the highest rating (five-star)
are shown in the second and third column. We also calculate the number of users who have posted the reviews
in the eighth column.

Second, we compare the similarity of the reviews from different users
using exact text matching. If we find that reviews from different users are exactly the same,
we classify such reviews as repeated reviews and log the number in the fourth column.
The users who have posted repeated reviews are classified as fake reviewers correspondingly (shown
in the ninth column). 

Third, we further mark all the reviews from fake reviewers
as fake reviews, which is shown in the fifth column. This step is added because the criteria used to determine repeated reviews is
too strict (exact text matching), and hence may have missed reviews with only little changes,
e.g., from the sentence \emph{``This is really a good app''} to \emph{``This is really an excellent app''}.
By adding all the reviews from users who have posted fake
reviews, we could cover the reviews that may be otherwise missed in the previous step. 

At last, we calculate the percentage of fake reviews and users in the sixth and last column.
We also calculate the percentage of five-star ratings of fake reviews (the seventh column).

The percentage of fake reviews are surprisingly high, where over $90\%$ of reviews in $10$ families are fake and more than $95\%$ ($96.70\%$) of the user ratings in the fake reviews are five-star, demonstrating that the ranking system is actively manipulated by the publishers of those FD apps.

\vspace{-0.1in}
\subsection{Advertising Networks}
\label{subsec:adnetwork}
Previous research~\cite{maliciousad} has revealed that 
mobile malware could be distributed through mobile advertising
networks. In this study, we perform an initial investigation to check
whether FD apps have been distributed through this channel.
Our study leverages a third-party online service AppGrowing~\cite{appgrowing}
to collect the corresponding data. In particular, given an app, AppGrowing provides a report containing whether this app has been distributed through an advertising network,
and if so which networks have been involved. Note that, their data is through \emph{sampling}
the traffics of advertisement SDKs and thus is not complete. Nevertheless, the data
still provide some insights of the distribution of FD apps.

Based on the three-month data from October to December 2017,
the following advertising networks have been involved in the distribution of FD apps: 
the Cheetah Ad~\cite{CheetahAds}, IntelligentTui~\cite{TuiQQ},
Tencent Social Ad~\cite{gdt} and Baidu Ad~\cite{baiduad}. 
The first one belongs to the Cheetah Mobile~\cite{Cheetah}, a company listed in New York Stock Exchange,
and the second and third advertising network belong to Tencent~\cite{Tencent}, the producer of QQ and WeChat and
one of the largest Internet and technology companies in the world. 
The last one belongs to Baidu~\cite{Baidu}, the largest searching engine and one of the biggest mobile advertising networks in China.
Table~\ref{table:adnetwork}
shows the $26$ apps we monitored and the advertising networks that have been leveraged to distribute those apps.
In the table, we use
the following abbreviations CH, IT, TD and BD to denote the 
Cheetah Ad, IntelligentTui, Tencent Social Ad and Baidu Ad, respectively.

\begin{table}[t]
\newcommand{\tabincell}[2]{\begin{tabular}{@{}#1@{}}#2\end{tabular}}
\caption{The package name of FD apps and their corresponding distributing advertising networks.}
\centering
\resizebox{1\linewidth}{!}{
\begin{tabular}{c|c||c|c}
\toprule
package name    &   AD Networks   &  Package Name  & AD Networks \\ \midrule
com.hzsj.kya                &  CH             &  com.aahl.zrl           & BD      \\
com.huizheng.tcyyhz         &  IT, CH         &  com.huizheng.kya       & IT, TE  \\   
com.huizheng.dsyyh          &  IT, CH         &  com.mimivip.missyou    & TE   \\  
com.huizheng.lasq           &  IT, CH, TE     &  com.hzsj.bdya          & IT  \\  
com.youyuan.yyhl            &  IT             &  com.solo.peanut        & IT \\  
com.hzsj.dsjy               &  IT, TE         &  com.huizheng.tcxax     & IT  \\  
com.dllingshang.tcjy        &  IT             &  com.yuanju.night       & IT \\  
com.futuredo.quickdate      &  CH             &  com.lingai.gaybar      & IT \\  
com.youyuan.yhb             &  IT, TE         &  com.meimei.yulove      & IT \\  
com.tanliani                &  CH             &  com.dljh.fjlxy         & IT \\  
com.youyuan.lrxq            & IT                &  com.huizheng.jrtt   & IT \\  
com.dlkuaidu.tcayh          & IT                &  com.prd.tosipai    & IT \\  
com.lingshang.ls            & IT                &  com.hzsj.zxzdxz    &BD \\  
\bottomrule
\end{tabular}
}
\label{table:adnetwork}
\end{table}



\section{User Impact Analysis}
\label{sec:impact}

In this section, we analyze the user impact of the FD apps from the following three aspects.
First, we measure the upper- and lower-bound of the number of victims. In particular, we first measure
the downloads distribution of these apps, which could be used to calculate the upper-bound of the number of victims. 
Then we crawl the negative comments of these apps from app markets, which could be used to estimate the
lower-bound of the number of victims, since these negative comments are likely to be posted by real victims.
Second, we estimate the overall revenue of these apps based on several reports that disclose the revenue model of these FD apps.
Third, we upload all these apps to VirusTotal to understand how many of them could be flagged by existing anti-virus engines.



\subsection{Number of Victims Estimation}

\begin{figure}[t]
 \centering
 \includegraphics[width=3in]{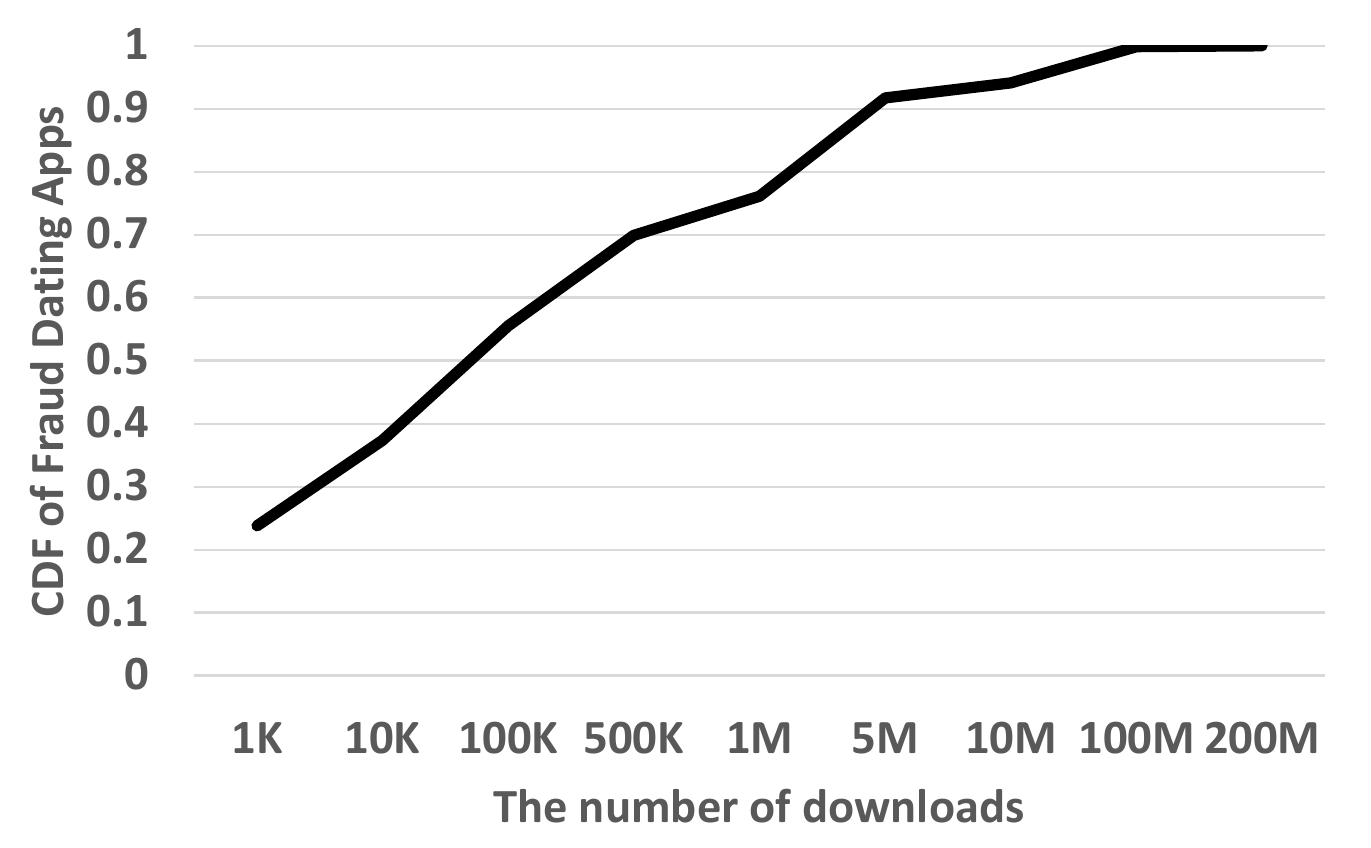}\\
 \caption{The cumulative distribution of the number of app downloads.}
 \label{Figure:downloadsCDF}
 \end{figure}

\smallskip
\noindent{\bf Downloads Analysis: the Upper-bound}\tab
We first analyze the number of accumulated downloads for these apps crawled from 10 app markets. Figure~\ref{Figure:downloadsCDF} shows the distribution. Surprisingly, around 50\% of apps have been downloaded more than 100K times, and roughly 25\% of them have the number of downloads over 1 million. App \code{cn.feichengwuyue} has the most number of downloads (143 million).

We then examine the downloads distribution across families. As shown in Table~\ref{table:userimpact}, there are 6 families that have achieved more than 100 million accumulated downloads, in which the family \code{Youyuan} has accumulated downloads of 784 million. The total number of downloads for these 967 apps is surprisingly high, which
has achieved 2.4 billion. Each app has an average of 2.5 million of downloads. 

\begin{table*}[h]
\newcommand{\tabincell}[2]{\begin{tabular}{@{}#1@{}}#2\end{tabular}}
\caption{User impact analysis.}
\label{table:userimpact}
\centering
\small
\resizebox{\linewidth}{!}{
\begin{tabular}{llllllll}
\toprule
Family & \#DLs & \#\tabincell{l}{Avg DLs} & \#\tabincell{l}{Negavive\\Reviews}  & \tabincell{l}{Est \# of\\Payment} & \tabincell{l}{Est Avg \# of\\Payment} & \tabincell{l}{Profit Est (US \$)} & \tabincell{l}{Avg Profit Est (US \$)} \\ \midrule
Youyuan  & 784.8M & 1582K & 22899 & 7.8M-78.4M & 16K-158K& 73.1M-731.1M & 147K-1474K  \\
Appforwhom  & 175.9M & 2513K & 2637 & 1.7M-17.5M & 25K-251K & 13.9M-138.9M & 198K-1984K \\
Youairen & 684.7M & 4890K & 6393 & 6.8M-68.5M & 49K-489K & 54.1M-540.5M & 386K-3861K  \\
Yueaiapp  & 35.4M & 1769K & 1186 & 0.4M-3.5M & 18K-177K & 3.8M-38M & 190K-1899K \\
Tanliani  & 189.2M & 7275K & 2161 & 1.9M-18.9M & 73K-728K & 8.1M-80.6M & 310K-3102K \\
Wmlover  & 177.5M & 5548K & 2490 & 1.8M-17.8M & 55K-555K & 14M-140.2M & 438K-4380K  \\
Tongchengsupei  & 5.3M & 126K & 136 & 53K-531K & 1K-13K & 0.6M-5.8M & 14K-138K \\
Jiangaijiaoyou  & 6.2M & 478K & 1210 & 62K-622K & 5K-48K  & 0.7M-6.8M & 52K-521K  \\
Aiaihunlian     & 4.9M & 486K & 358 & 49K-486K & 5K-49K & 0.5M-5.3M & 53K-529K  \\
Sipuhaiwei      & 65.2M & 5013K & 570 & 0.7M-6.5M & 50K-501K & 5.1M-51.5M & 396K-3958K \\
Yuanfenba      & 9.1M & 831K & 591 & 91K-914K & 8K-83K  & 0.7M-7.2M & 66K-656K  \\
Yuanlai         & 13.7M & 1145K & 1400 & 0.1M-1.4M  & 11K-115K & 0.7M-6.5M & 54K-543K \\
Qianshoulian    & 92.4M & 6163K & 1055 & 1M-9.2M & 62K-616K & 7.3M-73M & 487K-4866K \\
Erwanshenghuo   & 11.7M & 2937K & 55 & 0.1M-1.2M & 29K-294K  & 0.9M-9.3M & 232K-2319K \\
Zlewx           & 2.5M & 496K & 40 & 25K-248K & 5K-50K & 0.1M-1.2M & 23K-235K \\
Xiangyue        & 143.4M & 5735K & 259 & 1.4M-14.3M & 57K-574K & 10.9M-108.7M & 435K-4347K \\
99Paoyuan       & 6.3M & 1255K & 481 & 63K-628K & 13K-126K & 1M-9.7M & 194K-1943K \\
Qiaiapp         & 4.7M & 365K & 389 & 47K-475K & 4K-37K & 0.4M-3.7M & 29K-288K \\
Meiguihunlian   & 0.8M & 412K & 89 & 8K-82.4K & 4K-41K & 65K-651K & 33K-325K \\
Jucomic         & 10.9M & 1357K & 72 & 0.1M-1M & 14K-136K & 0.5M-5.1M & 64K-643K \\
Ailiaoba        & 1.5M & 759K & 29 & 15K-152K & 8K-76K & 0.1M-1.2M & 60K-599K \\
Michun          & 4.1M & 1374K & 252 & 41K-412K & 14K-137K & 0.2M-2M & 65K-651K \\
\midrule
Overall & 2.4B & 2.5M & 44752 & 24.3M-243M & 525K-5251K  &  196.7M-1966.9M & 3.9M-39.3M\\
\bottomrule
\end{tabular}
}
\end{table*}

\begin{figure}[t]
 \centering
 \includegraphics[width=3in]{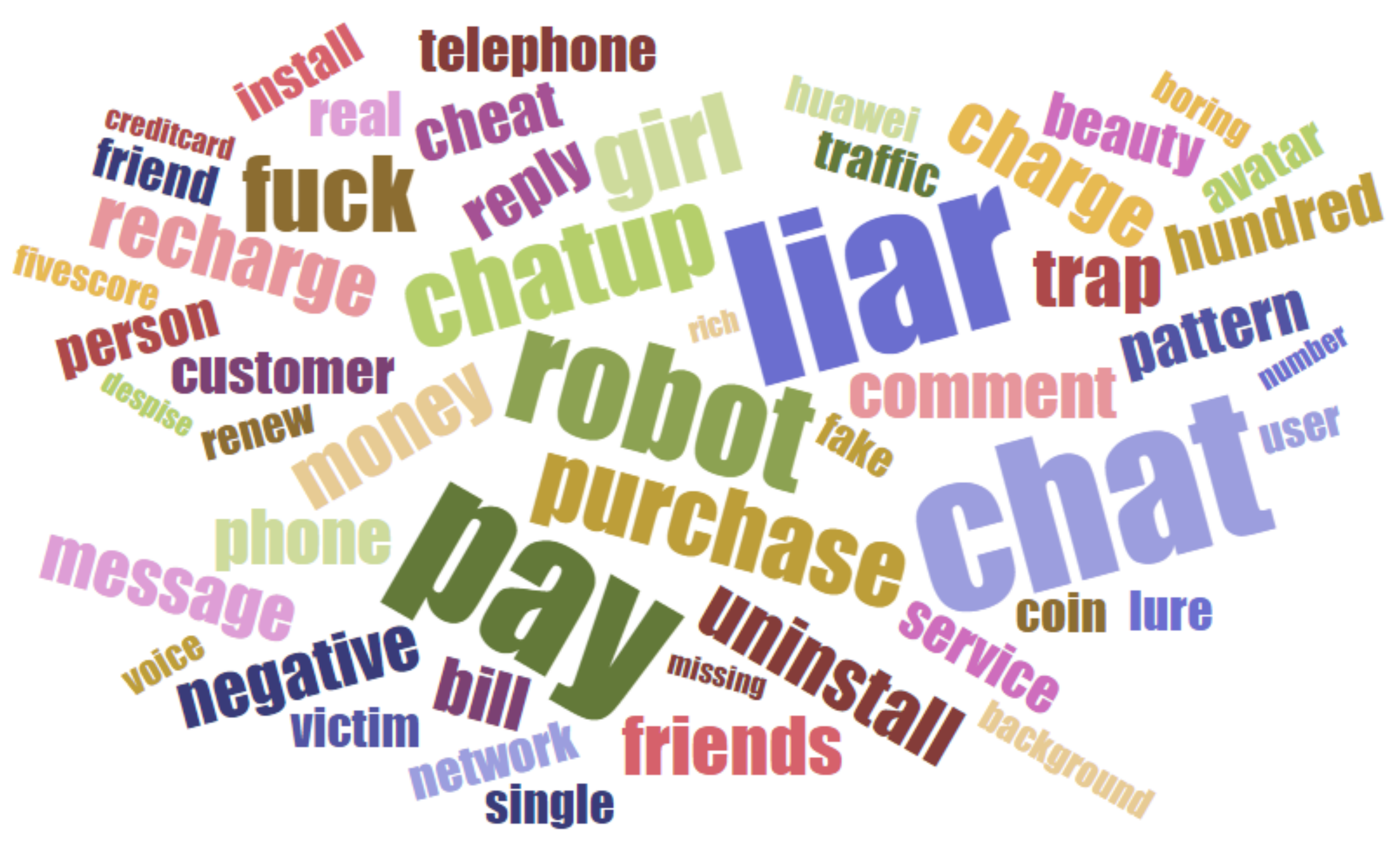}\\
 \caption{Top words in the 44,752 negative reviews.}
 \label{Figure:negativeReview}
 \end{figure}

\smallskip
\noindent{\bf Negative Review Analysis: the Lower-bound}\tab
Although we have shown that the positive reviews of these apps are mainly manipulated by ranking fraud techniques, the negative reviews
are usually about the complaints of victim users. We have analyzed the negative reviews of these apps. Figure~\ref{Figure:negativeReview}
shows the word cloud for the negative reviews. Almost all of the users complain that they have been cheated to purchase the premium services. 

Thus we measure the number of victims based on the negative comments.
As shown in Table~\ref{table:userimpact}, we have collected $44,752$ negative reviews in total,
and the family \code{Youyuan} has occupied more than half of the negative comments.
This result could be used to estimate the lower-bound of the victims.

\subsection{Payment Method Analysis and Profit Estimation}

\subsubsection{Price and Payment Method Analysis}

\begin{figure}[t]
 \centering
 \includegraphics[width=3in]{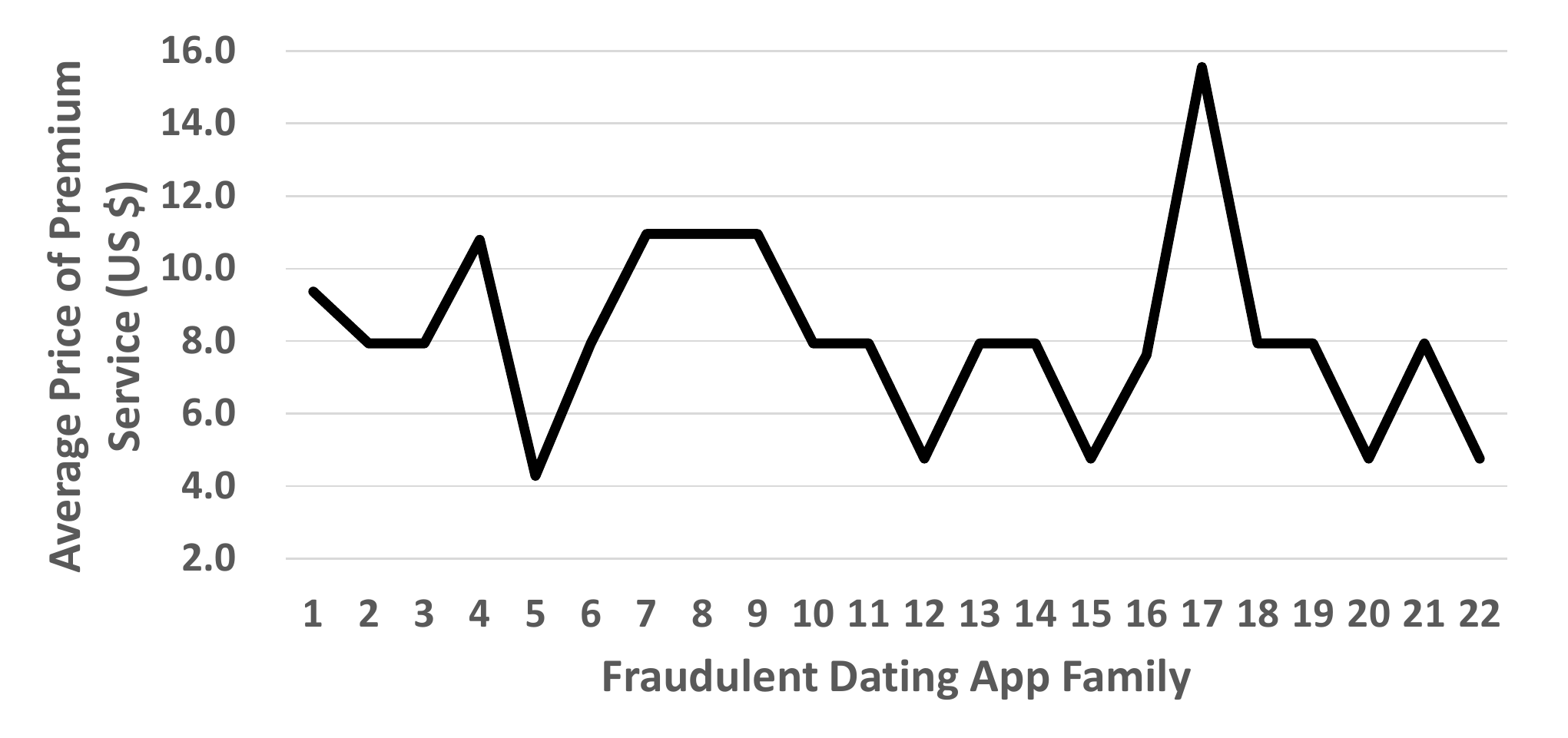}\\
 \caption{The average subscription price for premium services.}
 \label{Figure:price}
 \end{figure}

For each family, we randomly choose three apps to analyze the price of the premium services
they offered. As shown in Figure~\ref{Figure:price}, the average price varies from 5 US dollars to 15 US dollars.
We further analyze the payment method for each family. All the families support Alipay and WeChatPay, roughly $80\%$ of them also embed the Union Pay service, while 30\% of them provide the functionalities to send SMS to premium numbers.  

 \subsubsection{Payment Identifier Analysis} 
 
 
To use WeChatPay, the merchant should define three necessary parameters: \code{appid}, \code{mch\_id} and \code{secret key}. The appid and mch\_id are usually hard-coded in the app by app developers, which are used as the identification of the merchant. 
Note that the appid is an 18-byte string with the prefix wx, the mch\_id is a 10-byte digit string. Thus we first locate eligible strings in the decompiled code and then query the WeChatPay Web API to check whether we find the correct strings. At last, we have identified 232 unique WeChatPay identifiers (appid). With further analysis, we found that \emph{one developer signature usually corresponds to several payment identifiers, while one payment identifier always corresponds to one company name (the distributor)}. This finding once again provides evidence suggesting that app developers might sell FD apps to different distributors.

\subsubsection{Profit Estimation of FD Apps}

It is non-trivial for us to estimate the profit of these FD apps. Although we found several apps have the vulnerabilities (e.g., leaking their WeChat Payment security key) that could be exploited, we do not resort to exploit these apps to collect the unpublished revenue data due to ethical consideration. Several reports\footnote{http://www.thatsmags.com/shenzhen/post/21999/sexy-girl-bots-scam-1-billion-from-dating-app-users-in-china\\
https://www.bbc.com/news/blogs-news-from-elsewhere-42609353\\
http://www.marketing-interactive.com/12-chinese-dating-apps-close-down-after-women-found-to-be-robots/\\
http://www.cngold.com.cn/zjs/20180109d1897n202734126.html\\
http://www.lc123.net/xw/tp/2018-01-12/852867.html\\
http://cj.sina.com.cn/article/detail/1642467340/548172} have disclosed some information related with the revenue of this kind of apps. For example, it is reported that the Chinese polices have uncovered a case of fraudulent dating app at the early 2018, and the app was reported to have the revenue of more than 100 million US dollars one year. 
Thus we resort to these reports, and based on the real-world cases mentioned in the reports, we estimate the payment rate of these apps varies between 1\% to 10\% of the download number on average. As shown in Table~\ref{table:userimpact}, the accumulated revenue estimated for all the FD apps in this paper could be around 200 million US Dollars to 2 Billion US Dollars, and each app has an estimated revenue around 4 Million dollars to 40 million dollars, which is in line with the referred reports.




\subsection{Detection Results of VirusTotal}

We upload all the identified FD apps to VirusTotal to explore how many of them could be flagged by existing anti-virus engines. 
Surprisingly, more than half of them are labeled as malware by less than 1 anti-virus engine, meaning that most anti-virus engines are not able to flag those FD apps.
Only 5\% of these apps are flagged by more than 10 anti-virus engines. This result suggests that the FD apps cannot be sufficiently identified by existing anti-virus engines. 

We then analyze the distribution of malware families labeled by VirusTotal, as shown in Table~\ref{table:labelVT}. 
It is interesting to see that although roughly 700 APKs (18.7\%) are labeled as LoveFraud (PUA) by at least one engine, more than 80\% of the apps in our dataset are not identified as LoveFraud, even if they share the same behaviors and belong to the same families. 

 \begin{figure}[t]
 \centering
 \includegraphics[width=3in]{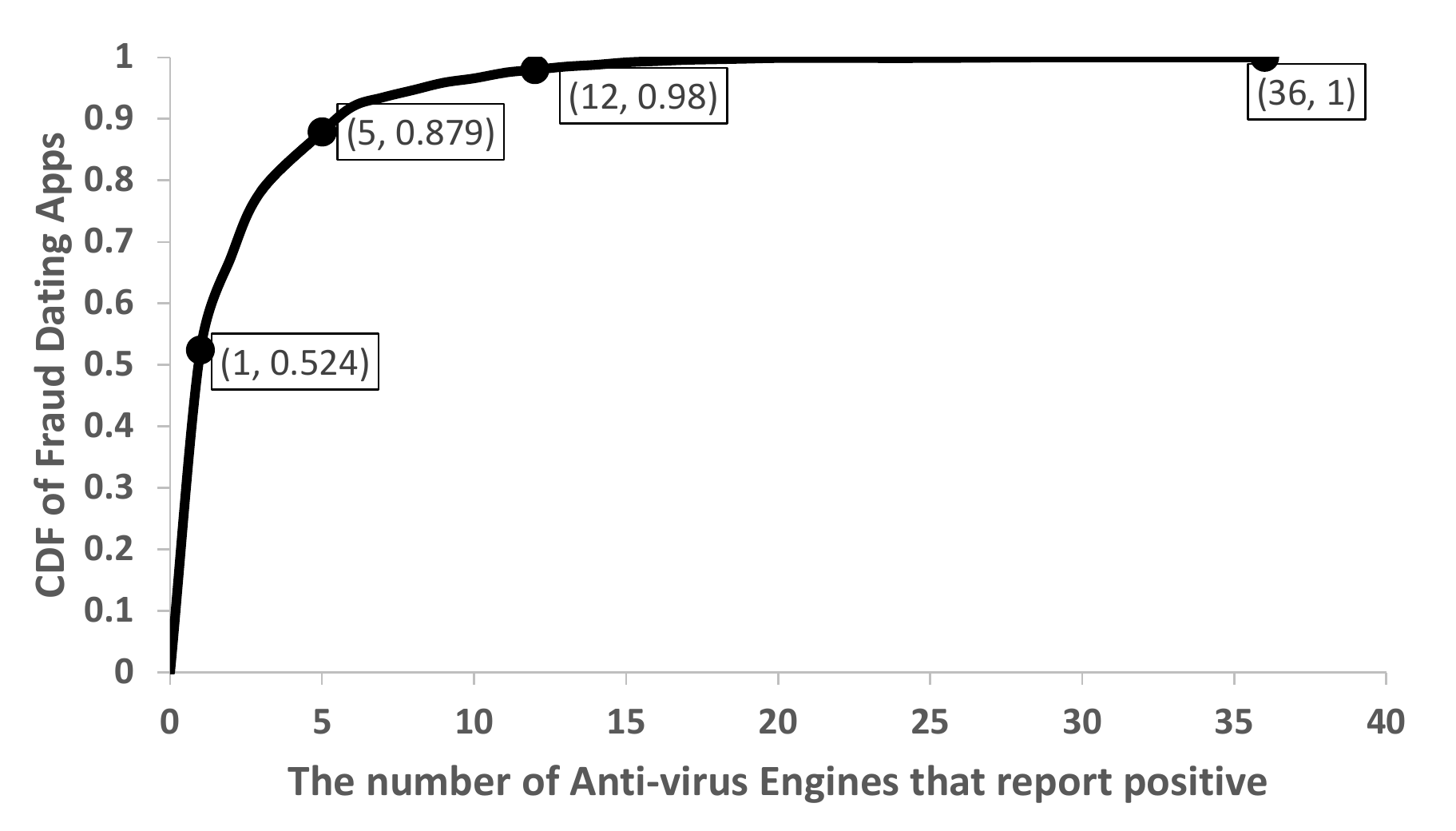}\\
 \caption{The distribution of VirusTotal Flags.}
 \label{Figure:virustotal}
 \end{figure}

\begin{table}[t]
\newcommand{\tabincell}[2]{\begin{tabular}{@{}#1@{}}#2\end{tabular}}
\centering
\small
\caption{The distribution of malware samples labeled by VirusTotal.}
\begin{tabular}{lcc}
\toprule
Malware Family & Count & Percent \\ \midrule
Trojan App   & 1138 & 30.7\% \\
Android-PUP & 813 & 21.9\% \\
LoveFraud (PUA) & 694 & 18.7\%\\
Adware & 377 & 10.2\% \\
LustFishingMoney & 259 & 7.0\% \\
Riskware App & 184 & 5.0\% \\
\bottomrule
\end{tabular}
\label{table:labelVT}
\end{table}

\section{Discussions}
\label{sec:discussion}


\subsection{Implication}
Besides showing the fact that there are many FD apps, our paper also delivers the following implications:

\noindent{\bf New approaches to detect FD apps.}\tab As demonstrated experimentally, FD apps cannot be sufficiently identified by VirusTotal, showing that our community needs to introduce new automated/semi-automated approaches to detect them. The various characteristics of FD apps summarized in this work could be helpful to create such detectors. For example, one malicious developer signature usually corresponds to several company names, which could be helpful in identifying suspicious apps in the markets. Besides, the detection results of FD apps could be used to help policies and regulators to identify the fraud rings behind-the-scenes.


\noindent{\bf A new and yet uncovered possible business model for developing and distributing malware.}\tab Empirically, we found that many FD apps share similar code implementation (e.g., unlikely be implemented independently) while being released with different package names by different companies. We hence hypothesize that these apps are bought (cloned) from some app developers by these so-called distributors (implemented once and sold many times). This sheds light on the possible business model of other kinds of malware (e.g., ransomware), though further investigations are expected.


\noindent{\bf Ranking fraud in app markets.}\tab FD apps use ASO methods (e.g., fake positive reviews) and various channels (app markets and ad networks) for app promotion and distribution, which could offer insights for general malware detection. Moreover, it raises the implication that market operators need to apply effective means to detect and hence avoid ranking fraud.

\subsection{Ethical Consideration}

Indeed, any data collected from real users need to be carefully processed. We take a series of steps to preserve the privacy of (possible) involved users/malicious developers in our data set. First, all raw data collected for this study are open to public, we do not resort to exploit the apps to collect the unpublished data (e.g., the revenue data), even though we found several apps have the vulnerabilities could be exploited. Second, we do not store all the account profiles we crawled from the apps after our experiment, even though almost all of them are fake. Third, all the user avatars we listed in the paper are convinced to be fake profiles and they could be found on the public INTERNET, which we believe do not violate the privacy of them.

\subsection{Limitation}

First, the method used to detect the FD apps is conservative and may
miss some of them. For instance, we use the keywords and embedded in-app purchase libraries
to find candidate apps. Though this method leads us to the discovery of
$23,546$ candidate apps, the list of keywords and in-app purchase libraries may not be complete
and could introduce false negatives. 
In addition, we use the heuristic to find the fake reviews. Specifically, we use
the exact text matching to find the repeated reviews. This may miss the reviews
that have same meanings but in different texts.


Second, we find that FD apps are distributed through app markets
and advertising networks. These include the app markets of leading phone vendors, e.g.,
Huawei, and advertising networks from world-class Internet
companies, e.g., Baidu and Tencent. The detection result of VirusTotal
shows that most anti-virus engines cannot detect
these apps. These worrisome facts urge a more effective detection schema of these apps,
and a better vetting process of apps in app markets and advertising networks. 



Our study is mainly focused on the apps in the Chinese app markets.
Most of the fraudulent dating apps analyzed in this paper are targeting users in China, 
possibly due to the biased region distribution of apps in our dataset. 
However, we believe such kind of apps may exist in many other countries and in other languages as well, especially the places where app vetting is not strictly enforced when they are uploaded to an app market.
In our future work, we plan to extend our crawler to download more apps from app markets in other countries.




\section{Related Work}
\label{sec:related}

This paper is motivated by the work of Caballero et al.~\cite{caballero2011measuring} and of Thomas~\cite{thomas2016investigating}, who have investigated the so-called ``underground economy'' associated with malicious apps (or unwanted software).
FD apps fall into the same research line. To the best of our knowledge, the ecosystem of FD apps has not yet been investigated.
Nevertheless, various studies have explored the general fraudulent behaviors in the mobile app ecosystem as well as the security aspect of dating apps.



\subsection{Fraudulent Behaviors}
Fraudulent behaviors have been widely explored in the mobile app ecosystem~\cite{yeh2015fraud,de2017android,liu2016you,li2017static,dong2017fraudroid,dong2018mobile}.
The most common issue is ad fraud, where a miscreant's code fetches ads without displaying them to the user or ``clicks'' ads automatically~\cite{crussell2014madfraud,blizard2012click,stevens2012investigating,liu2014decaf,cho2016combating}.
For example, Crussell et al.~\cite{crussell2014madfraud} have revealed two fraudulent ad behaviors: (1) ads are requested by apps that are running in the background and (2) ads are clicked without user interaction (also known as click frauds~\cite{cho2015empirical}).
More recently, Dong et al.~\cite{dong2017fraudroid} reveal seven types of ad frauds and further demonstrate that such ad fraudulent apps are also likely to violate the policy of app markets, resulting in risks to be removed from app markets~\cite{dong2018mobile}.  
Besides ad fraud, there are also other types of frauds disclosed by several researchers.
For example,  Liu et al.~\cite{liu2016you} have explored \emph{usage fraud}, which is invented to boost usage statistics on third-party analytics like Google Analytics, resulting in inaccurate numbers that could eventually fool investors to make wrong decisions. 
Xie et al.~\cite{xie2015appwatcher} have analyzed the review fraud in mobile apps and found that some mobile app developers turn to the underground market to buy positive reviews. We also found in this paper that FD apps use fake positive reviews for app promotion.
Our work is focusing on fraudulent dating Android apps and has revealed a new type of fraudulent behaviors, which have not been systemically studied.

\subsection{Security and Privacy in Dating Apps/Online Dating Website}

Dating apps have raised security and privacy-related concerns in recent years~\cite{shetty2017you,farnden2015privacy,spensky2016sok,imgraben2014always,argyros2017evaluating}.
As shown by Shetty et al.~\cite{shetty2017you}, mobile dating apps are potentially vulnerable to security risks.
For example, they have demonstrated that it is quite trivial to conduct a man-in-the-middle attack against most dating apps, resulting in private data leaks of app users.
Similarly, Hoang et al.~\cite{hoang2017your} argue that trilateration threatens location privacy of users of location-based mobile apps.
They demonstrate that it is possible for an adversary to identify the location of an individual when she/he is using dating apps, even under the situation where location-hiding features are enabled. Similar findings have also been reported by Carman et al.~\cite{carman2016tinder}, who have empirically presented their experiments on a popular dating app called Tinder. In addition to leaking location information directly, certain sensitive information (e.g., nearly usernames, profile pictures, messages, etc.) can also be recovered from user's devices based on the residual data generated by dating apps~\cite{farnden2015privacy}. 
Our work is not towards the potential security and privacy concerns of legitimate dating apps, but revealing a new type of malicious dating apps, i.e., fraudulent dating apps.

Moreover, we want to clarify that, the fraudulent behaviors of these apps
are different from the well-known \emph{romance scam}~\cite{romancescam, onlinescam1, rege2009s, whitty2012online}. 
In particular, the fraudulent acts in romance scam are usually with the involvement of real persons,
who are communicating with victims through phones, emails and try to access to
victims' money or bank account. However, in FD apps,
the chatbots instead of real persons, are communicating with victims.
The main purpose is to lure the victim into buying premium service, not the
financial information.
Though, we find there are still some common aspects between them.
For instance, seductive account profile avatars are used to attract victims in both
romance scam and FD apps.

\section{Conclusion}
\label{sec:conclusion}
In this work, we perform a systematic study of fraudulent dating apps,
including its characteristics, business model, distribution networks
and their impact on affected users. Our research has observed various findings that are previously unknown to the community. Due to the financial loss to victims and the fact that current anti-virus engines cannot detect most of these apps, we argue that an effective solution should be proposed to detect such apps or block the distribution of these apps in the first place to protect users.

\bibliographystyle{IEEEtran}
\bibliography{fraudapp}

\begin{thebibliography}{10}
\providecommand{\url}[1]{#1}
\csname url@samestyle\endcsname
\providecommand{\newblock}{\relax}
\providecommand{\bibinfo}[2]{#2}
\providecommand{\BIBentrySTDinterwordspacing}{\spaceskip=0pt\relax}
\providecommand{\BIBentryALTinterwordstretchfactor}{4}
\providecommand{\BIBentryALTinterwordspacing}{\spaceskip=\fontdimen2\font plus
\BIBentryALTinterwordstretchfactor\fontdimen3\font minus
  \fontdimen4\font\relax}
\providecommand{\BIBforeignlanguage}[2]{{%
\expandafter\ifx\csname l@#1\endcsname\relax
\typeout{** WARNING: IEEEtran.bst: No hyphenation pattern has been}%
\typeout{** loaded for the language `#1'. Using the pattern for}%
\typeout{** the default language instead.}%
\else
\language=\csname l@#1\endcsname
\fi
#2}}
\providecommand{\BIBdecl}{\relax}
\BIBdecl

\bibitem{PiggyApp}
W.~Zhou, Y.~Zhou, M.~Grace, X.~Jiang, and S.~Zou, ``{Fast, Scalable Detection
  of "Piggybacked" Mobile Applications},'' in \emph{{Proceedings of the 3rd ACM
  Conference on Data and Application Security and Privacy}}, 2013.

\bibitem{Sok}
Y.~Zhou and X.~Jiang, ``{Dissecting Android Malware: Characterization and
  Evolution},'' in \emph{Proceedings of the 33rd IEEE Symposium on Security and
  Privacy}, 2012.

\bibitem{libradar}
Z.~Ma, H.~Wang, Y.~Guo, and X.~Chen, ``Libradar: fast and accurate detection of
  third-party libraries in android apps,'' in \emph{Proceedings of the 38th
  International Conference on Software Engineering Companion}, 2016, pp.
  653--656.

\bibitem{alipay}
``{AliPay},'' 2018, \url{https://www.alipay.com}.

\bibitem{wexpay}
``{WeChatPay},'' 2018, \url{https://pay.weixin.qq.com/index.php/core/home/}.

\bibitem{FSquaDRA2}
O.~Gadyatskaya, A.-L. Lezza, and Y.~Zhauniarovich, ``{Evaluation of
  Resource-based App Repackaging Detection in Android},'' in \emph{Proceedings
  of the 21st Nordic Conference on Secure IT Systems}, ser. NordSec 2016, 2016,
  pp. 135--151.

\bibitem{wukong}
H.~Wang, Y.~Guo, Z.~Ma, and X.~Chen, ``Wukong: A scalable and accurate
  two-phase approach to android app clone detection,'' in \emph{Proceedings of
  the 2015 International Symposium on Software Testing and Analysis}, 2015, pp.
  71--82.

\bibitem{droidbot}
Y.~Li, Z.~Yang, Y.~Guo, and X.~Chen, ``Droidbot: a lightweight ui-guided test
  input generator for android,'' in \emph{Software Engineering Companion
  (ICSE-C), 2017 IEEE/ACM 39th International Conference on}.\hskip 1em plus
  0.5em minus 0.4em\relax IEEE, 2017, pp. 23--26.

\bibitem{romancescam}
``{Romance Scams: Online Imposters Break Hearts and Bank Accounts},'' 2018,
  \url{https://www.fbi.gov/news/stories/romance-scams}.

\bibitem{dupdetector}
``{Dup Detector},'' 2018, \url{https://www.keronsoft.com/dupdetector.html}.

\bibitem{legalperson}
``{Legal person},'' 2018, \url{https://en.wikipedia.org/wiki/Legal_person}.

\bibitem{appgrowing}
``{appgrowing},'' 2018, \url{https://appgrowing.cn}.

\bibitem{rankingfraud}
H.~Zhu, H.~Xiong, Y.~Ge, and E.~Chen, ``{Discovery of Ranking Fraud for Mobile
  Apps},'' in \emph{{IEEE Transactions on Knowledge and Data Engineering}},
  2015.

\bibitem{maliciousad}
V.~Rastogi, R.~Shao, Y.~Chen, X.~Pan, S.~Zou, and R.~Riley, ``{Are these Ads
  Safe: Detecting Hidden Attacks through the Mobile App-Web Interfaces},'' in
  \emph{{Proceedings of the Network and Distributed System Security
  Symposium}}, 2016.

\bibitem{CheetahAds}
``{Cheetah Ad},'' 2018, \url{http://ads.cmcm.com}.

\bibitem{TuiQQ}
``{IntelligentTui},'' 2018, \url{https://tui.qq.com}.

\bibitem{gdt}
``{Tencent Social Ad},'' 2018, \url{http://ads.tencent.com}.

\bibitem{baiduad}
``{Baidu Ad},'' 2018, \url{http://mssp.baidu.com/home}.

\bibitem{Cheetah}
``{Cheetah Mobile},'' 2018, \url{https://en.wikipedia.org/wiki/Cheetah_Mobile}.

\bibitem{Tencent}
``{Tencent},'' 2018, \url{https://en.wikipedia.org/wiki/Tencent}.

\bibitem{Baidu}
``{Baidu},'' 2018, \url{https://en.wikipedia.org/wiki/Baidu}.

\bibitem{caballero2011measuring}
J.~Caballero, C.~Grier, C.~Kreibich, and V.~Paxson, ``Measuring
  pay-per-install: the commoditization of malware distribution.'' in
  \emph{Usenix security symposium}, 2011, pp. 13--13.

\bibitem{thomas2016investigating}
K.~Thomas, J.~A.~E. Crespo, R.~Rasti, J.~M. Picod, C.~Phillips, M.-A. Decoste,
  C.~Sharp, F.~Tirelo, A.~Tofigh, M.-A. Courteau \emph{et~al.}, ``Investigating
  commercial pay-per-install and the distribution of unwanted software.'' in
  \emph{USENIX Security Symposium}, 2016, pp. 721--739.

\bibitem{yeh2015fraud}
K.-H. Yeh, N.-W. Lo, L.-C. Chen, and P.-H. Lin, ``A fraud detection system for
  real-time messaging communication on android facebook messenger,'' in
  \emph{Consumer Electronics (GCCE), 2015 IEEE 4th Global Conference on}.\hskip
  1em plus 0.5em minus 0.4em\relax IEEE, 2015, pp. 361--363.

\bibitem{de2017android}
S.~de~los Santos, A.~Guzm{\'a}n, and C.~Torrano, ``Android malware pattern
  recognition for fraud detection and attribution: A case study,''
  \emph{Encyclopedia of Social Network Analysis and Mining}, pp. 1--9, 2017.

\bibitem{liu2016you}
W.~Liu, Y.~Zhang, Z.~Li, and H.~Duan, ``What you see isn't always what you get:
  A measurement study of usage fraud on android apps,'' in \emph{Proceedings of
  the 6th Workshop on Security and Privacy in Smartphones and Mobile
  Devices}.\hskip 1em plus 0.5em minus 0.4em\relax ACM, 2016, pp. 23--32.

\bibitem{li2017static}
L.~Li, T.~F. Bissyand{\'e}, M.~Papadakis, S.~Rasthofer, A.~Bartel, D.~Octeau,
  J.~Klein, and Y.~Le~Traon, ``Static analysis of android apps: A systematic
  literature review,'' \emph{Information and Software Technology}, 2017.

\bibitem{dong2017fraudroid}
F.~Dong, H.~Wang, Y.~Li, Y.~Guo, L.~Li, S.~Zhang, and G.~Xu, ``Fraudroid: An
  accurate and scalable approach to automated mobile ad fraud detection,''
  \emph{arXiv preprint arXiv:1709.01213}, 2017.

\bibitem{dong2018mobile}
F.~Dong, H.~Wang, L.~Li, Y.~Guo, G.~Xu, and S.~Zhang, ``How do mobile apps
  violate the behavioral policy of advertisement libraries?'' in
  \emph{Proceedings of the 19th International Workshop on Mobile Computing
  Systems \& Applications}.\hskip 1em plus 0.5em minus 0.4em\relax ACM, 2018,
  pp. 75--80.

\bibitem{crussell2014madfraud}
J.~Crussell, R.~Stevens, and H.~Chen, ``Madfraud: Investigating ad fraud in
  android applications,'' in \emph{Proceedings of the 12th annual international
  conference on Mobile systems, applications, and services}.\hskip 1em plus
  0.5em minus 0.4em\relax ACM, 2014, pp. 123--134.

\bibitem{blizard2012click}
T.~Blizard and N.~Livic, ``Click-fraud monetizing malware: A survey and case
  study,'' in \emph{Malicious and Unwanted Software (MALWARE), 2012 7th
  International Conference on}.\hskip 1em plus 0.5em minus 0.4em\relax IEEE,
  2012, pp. 67--72.

\bibitem{stevens2012investigating}
R.~Stevens, C.~Gibler, J.~Crussell, J.~Erickson, and H.~Chen, ``Investigating
  user privacy in android ad libraries,'' in \emph{Workshop on Mobile Security
  Technologies (MoST)}, vol.~10, 2012.

\bibitem{liu2014decaf}
B.~Liu, S.~Nath, R.~Govindan, and J.~Liu, ``Decaf: Detecting and characterizing
  ad fraud in mobile apps.'' in \emph{NSDI}, 2014, pp. 57--70.

\bibitem{cho2016combating}
G.~Cho, J.~Cho, Y.~Song, D.~Choi, and H.~Kim, ``Combating online fraud attacks
  in mobile-based advertising,'' \emph{EURASIP Journal on Information
  Security}, vol. 2016, no.~1, p.~2, 2016.

\bibitem{cho2015empirical}
G.~Cho, J.~Cho, Y.~Song, and H.~Kim, ``An empirical study of click fraud in
  mobile advertising networks,'' in \emph{Availability, Reliability and
  Security (ARES), 2015 10th International Conference on}.\hskip 1em plus 0.5em
  minus 0.4em\relax IEEE, 2015, pp. 382--388.

\bibitem{xie2015appwatcher}
Z.~Xie and S.~Zhu, ``Appwatcher: Unveiling the underground market of trading
  mobile app reviews,'' in \emph{Proceedings of the 8th ACM Conference on
  Security \& Privacy in Wireless and Mobile Networks}.\hskip 1em plus 0.5em
  minus 0.4em\relax ACM, 2015, p.~10.

\bibitem{shetty2017you}
R.~Shetty, G.~Grispos, and K.-K.~R. Choo, ``Are you dating danger? an
  interdisciplinary approach to evaluating the (in) security of android dating
  apps,'' \emph{IEEE Transactions on Sustainable Computing}, 2017.

\bibitem{farnden2015privacy}
J.~Farnden, B.~Martini, and K.-K.~R. Choo, ``Privacy risks in mobile dating
  apps,'' \emph{arXiv preprint arXiv:1505.02906}, 2015.

\bibitem{spensky2016sok}
C.~Spensky, J.~Stewart, A.~Yerukhimovich, R.~Shay, A.~Trachtenberg, R.~Housley,
  and R.~K. Cunningham, ``Sok: privacy on mobile devices--it’s complicated,''
  \emph{Proceedings on Privacy Enhancing Technologies}, vol. 2016, no.~3, pp.
  96--116, 2016.

\bibitem{imgraben2014always}
J.~Imgraben, A.~Engelbrecht, and K.-K.~R. Choo, ``Always connected, but are
  smart mobile users getting more security savvy? a survey of smart mobile
  device users,'' \emph{Behaviour \& Information Technology}, vol.~33, no.~12,
  pp. 1347--1360, 2014.

\bibitem{argyros2017evaluating}
G.~Argyros, T.~Petsios, S.~Sivakorn, A.~D. Keromytis, and J.~Polakis,
  ``Evaluating the privacy guarantees of location proximity services,''
  \emph{ACM Transactions on Privacy and Security (TOPS)}, vol.~19, no.~4,
  p.~12, 2017.

\bibitem{hoang2017your}
N.~P. Hoang, Y.~Asano, and M.~Yoshikawa, ``Your neighbors are my spies:
  Location and other privacy concerns in glbt-focused location-based dating
  applications,'' in \emph{Advanced Communication Technology (ICACT), 2017 19th
  International Conference on}.\hskip 1em plus 0.5em minus 0.4em\relax IEEE,
  2017, pp. 851--860.

\bibitem{carman2016tinder}
M.~Carman and K.-K.~R. Choo, ``Tinder me softly--how safe are you really on
  tinder?'' in \emph{International Conference on Security and Privacy in
  Communication Systems}.\hskip 1em plus 0.5em minus 0.4em\relax Springer,
  2016, pp. 271--286.

\bibitem{onlinescam1}
J.~Huang, G.~Stringhini, and P.~Yong, ``Quit playing games with my heart:
  Understanding online dating scams,'' in \emph{Proceedings of the 12th
  International Conference on Detection of Intrusions and Malware, and
  Vulnerability Assessment - Volume 9148}, ser. DIMVA 2015, 2015, pp. 216--236.

\bibitem{rege2009s}
A.~Rege, ``What's love got to do with it? exploring online dating scams and
  identity fraud,'' \emph{International Journal of Cyber Criminology}, vol.~3,
  no.~2, p. 494, 2009.

\bibitem{whitty2012online}
M.~T. Whitty and T.~Buchanan, ``The online romance scam: A serious
  cybercrime,'' \emph{CyberPsychology, Behavior, and Social Networking},
  vol.~15, no.~3, pp. 181--183, 2012.

\end{thebibliography}

\ifCLASSOPTIONcompsoc

\ifCLASSOPTIONcaptionsoff
  \newpage
\fi

\balance
\begin{IEEEbiography}{Yangyu Hu} was born in 1991. He received the bachelor degree in applied physics from Beijing University of Posts and Telecommunications, Beijing, in 2012. Currently, he is a Ph.D student in the school of Cyberspace Security at Beijing University of Posts and Telecommunications. His research interest includes mobile application security and privacy.
\end{IEEEbiography}
\vspace{-10 mm}

\begin{IEEEbiography}{Haoyu Wang} received his PhD degree from Peking University in 2016. He is currently an Assistant Professor at Beijing University of Posts and Telecommunications. His research interests lie at the intersection of mobile system, privacy and security, and program analysis.
\end{IEEEbiography}
\vspace{-10 mm}
\begin{IEEEbiography}{Yajin Zhou} earned his Ph.D. in Computer Science from North Carolina State University. He is currently a Professor of Zhejiang University. His research mainly focuses on smartphone and system security, i.e., identifying real-world threats and building practical solutions. Currently He is working on interesting smartphone security projects.
\end{IEEEbiography}
\vspace{-10 mm}
\begin{IEEEbiography}{Yao Guo} received his PhD in Computer Engineering from University of Massachusetts at Amherst in 2007. He is a full Professor in the Institute of Software of the School of Electronics Engineering and Computer Science at Peking University. He has served as Vice Chair of Computer Science since 2013. His general research interests include operating systems, mobile computing and applications, low-power design and software engineering.
\end{IEEEbiography}
\vspace{-10 mm}
\begin{IEEEbiography}{Li Li} is a Lecturer at the Faculty of Information Technology, Monash University. Prior to join Monash, he was a research associate in Software Engineering at the University of Luxembourg, where he obtained his PhD degree in 2016. His research interests are in the fields of Android security, static code analysis, and machine learning. Dr. Li received a Best Paper Award at the ERA track of IEEE SANER 2016 and the FOSS Impact Paper Award at MSR 2018.
\end{IEEEbiography}
\vspace{-10 mm}
\begin{IEEEbiography}{Bingxuan Luo} is a student at Beijing University of Posts and Telecommunications. She was a research assistant at Professor Haoyu Wang's group. Her research interest is mobile security.
\end{IEEEbiography}
\vspace{-10 mm}
\begin{IEEEbiography}{Fangren Xu} is a student of Rivermont Collegiate. He was a research assistant at Professor Haoyu Wang's group. His research interest is mobile security.
\end{IEEEbiography}
\vspace{-10 mm}


\end{document}